\documentclass[twocolumn,floats,prd,amssymb,floatfix,nofootinbib,balancelastpage,superscriptaddress,amsmath]{revtex4-1}

\pdfoutput=1

\usepackage{epsfig,units}
\usepackage{latexsym}
\usepackage{graphics,graphicx,color,amssymb,float,amsmath}
\usepackage{pstricks}
\usepackage{color,slashed}
\usepackage{mathrsfs}
\usepackage{xspace}
\usepackage{soul}
\usepackage[utf8]{inputenc}

\def\beqn{\begin{eqnarray}} 
\def\eeqn{\end{eqnarray}} 
\def\be{\begin{equation}}
\def\ee{\end{equation}}

\newcommand{\tax}{{\tilde{a}}}

\newcommand{\Dslash}[1]{#1 \llap{/\kern+1.2pt}}
\newcommand{\neu}{\widetilde{\chi}^0_1}

\newcommand{\cc}{\mathcal{C}}
\newcommand{\li}{\ell^i}
\newcommand{\vi}{\nu^i}

\newcommand{\gneudR}{g_{\widetilde d_R}}

\newcommand{\newc}{\newcommand}
\newc{\half}{\frac{1}{2}}
\newc{\Lam}{{\bf \Lambda}}
\newc{\ltau}{\lambda_\tau}
\newc{\lt}{\lambda_t}
\newc{\lb}{\lambda_b}
\newc{\kap}{{\bf \kappa}}
\newc{\lae}{{\Lam}_E}
\newc{\lad}{{\Lam}_D}
\newc{\lau}{{\Lam}_U}
\newc{\lame}[1]{{\Lam}_{E^{#1}}}
\newc{\lamhe}[1]{{\h}_{E^{#1}}}
\newc{\lamhed}[1]{{\h}_{E^{#1}}^\dagger}
\newc{\lamhd}[1]{{\h}_{D^{#1}}}
\newc{\lamhdd}[1]{{\h}_{D^{#1}}^\dagger}
\newc{\lamhu}[1]{{\h}_{U^{#1}}}
\newc{\lamhud}[1]{{\h}_{U^{#1}}^\dagger}
\newc{\lamd}[1]{{\Lam}_{D^{#1}}}
\newc{\lamu}[1]{{\Lam}_{U^{#1}}}
\newc{\lamet}[1]{{\Lam}_{E^{#1}}^T}
\newc{\lamdt}[1]{{\Lam}_{D^{#1}}^T}
\newc{\lamut}[1]{{\Lam}_{U^{#1}}^T}
\newc{\lames}[1]{{\Lam}_{E^{#1}}^*}
\newc{\lamds}[1]{{\Lam}_{D^{#1}}^*}
\newc{\lamus}[1]{{\Lam}_{U^{#1}}^*}
\newc{\lamed}[1]{{\Lam}_{E^{#1}}^\dagg}
\newc{\lamdd}[1]{{\Lam}_{D^{#1}}^\dagg}
\newc{\lamud}[1]{{\Lam}_{U^{#1}}^\dagg}
\newc{\lam}{{\bf \lambda}}
\newc{\lamp}{{\bf \lambda}^{\prime}}
\newc{\lampp}{{\bf \lambda}^{\prime\prime}}
\newc{\Y}{{\bf Y}}
\newc{\h}{{\bf h}}
\newc{\meee}{{{\rm {\bf  m}}_e}}
\newc{\mdee}{{{\rm {\bf  m}}_d}}
\newc{\myew}{{{\rm {\bf m}}_u}}
\newc{\ye}{{\Y}_E}
\newc{\he}{{\h}_E}
\newc{\hed}{{\h}_E^\dagger}
\newc{\yd}{{\Y}_D}
\newc{\hd}{{\h}_D}
\newc{\hdd}{{\h}_D^\dagger}
\newc{\yu}{{\Y}_U}
\newc{\hu}{{\h}_U}
\newc{\hud}{{\h}_U^\dagger}
\newc{\yes}{{\Y}_E^*}
\newc{\yds}{{\Y}_D^*}
\newc{\yus}{{\Y}_U^*}
\newc{\yet}{{\Y}_E^T}
\newc{\ydt}{{\Y}_D^T}
\newc{\yut}{{\Y}_U^T}
\newc{\yed}{{\Y}_E^\dagg}
\newc{\ydd}{{\Y}_D^\dagg}
\newc{\yud}{{\Y}_U^\dagg}
\newc{\dagg}{\dagger}

\newc{\fPQ}{f_{\rm{PQ}}}
\newc{\gsim}{\gtrsim}
\newc{\lsim}{\lessim}
\newc{\nonum}{\nonumber}

\newc{\hdtext}[1]{{\color{red} HD: #1}}
\newc{\htext}[1]{{\color{red} #1}}
\newc{\jtext}[1]{{\color{brown} #1}}
\newc{\bltext}[1]{{\color{blue}  #1}}
\newc{\datext}[1]{{\color{purple}  #1}}

\newc{\kpvertex}{\Phi_{\tax}H_2H_1}
\newc{\kppvertex}{\Phi_{\tax}L_iH_2}

\begin{document}
 \DeclareGraphicsExtensions{.pdf,.png,.gif,.jpg}

\title{R-Parity Violation and Light Neutralinos at SHiP and the LHC}

\author{Jordy de Vries}
\email{j.de.vries@fz-juelich.de}
\affiliation{Institute for Advanced Simulation, Institut f\"ur Kernphysik,
J\"ulich Center for Hadron Physics, 
Forschungszentrum J\"ulich, D-52425 J\"ulich, Germany}

\author{Herbi K. Dreiner}
\email{dreiner@uni--bonn.de}
\affiliation{Physikalisches Institut der Universit\"at Bonn, Bethe Center
for Theoretical Physics, \\ Nu{\ss}allee 12, 53115 Bonn, Germany}

\author{Daniel Schmeier}
\email{daschm@th.physik.uni--bonn.de}
\affiliation{Physikalisches Institut der Universit\"at Bonn, Bethe Center
for Theoretical Physics, \\ Nu{\ss}allee 12, 53115 Bonn, Germany}

 \begin{picture}(0,0)
  \put(0,0){BONN-TH-2015-12}
 \end{picture}

\begin{abstract}
We study the sensitivity of the proposed SHiP experiment to the $LQD$ operator in R-Parity violating 
supersymmetric theories. We focus on single neutralino production via rare meson decays and the 
observation of downstream neutralino decays into charged mesons inside the SHiP decay chamber. 
We provide a generic list of effective operators and decay width formulae for any $\lambda^\prime$ 
coupling and show the resulting expected SHiP sensitivity for a widespread list of benchmark scenarios 
via numerical simulations. We compare this sensitivity to expected limits from testing the same decay topology at the LHC with ATLAS.
\end{abstract}

\maketitle

\section{Introduction}
\label{sec:Introduction}
Supersymmetry \cite{Wess:1974tw,Nilles:1983ge,Martin:1997ns} is the unique extension of the external 
symmetries of the Standard Model of elementary particle physics (SM) with fermionic generators 
\cite{Haag:1974qh}. Supersymmetry is necessarily broken, in order to comply with the bounds from 
experimental searches. To solve the hierarchy problem \cite{Gildener:1976ai,Veltman:1980mj},
the masses of the supersymmetric partners of the SM fields should be lighter than 1-10\,TeV, with a clear 
preference for lighter fields, see for example \cite{Bechtle:2012zk,Baer:2012cf,CahillRowley:2012rv,
Ross:2012nr,Ghilencea:2013nxa}. All experimental searches for supersymmetric particles have hitherto 
been unsuccessful. The LHC sets limits on the strongly interacting sparticles, the squarks and gluinos, of 
order 1\,TeV, see for example \cite{Aad:2015wqa,Khachatryan:2015vra}. However the limits on the only 
electroweak interacting particles, such as the sleptons, the neutralinos and the charginos, are significantly
weaker \cite{Aad:2014vma,Khachatryan:2014mma}. In fact, as has been 
discussed in the literature there is currently \textit{no lower mass bound} on the lightest supersymmetric 
particle (LSP) neutralino, which is model 
independent \cite{Dreiner:2009ic,Agashe:2014kda}. The limits from LEP can easily be avoided, and in particular, 
a massless neutralino is still allowed, for sufficiently heavy selectrons and squarks \cite{Dreiner:2003wh}. We 
discuss this in more detail below, in Sec.~\ref{sec:light-neutralino}. 

We are here interested in the possibilities of searching for a light neutralino, up to masses of about 10 GeV. 
Already for a slepton mass of 150 GeV there is no sensitivity via the process $e^+e^-\to\tilde\chi^0_1\chi^0_1
\gamma$ from LEP data \cite{Choudhury:1999tn}, see also \cite{Dreiner:2006sb,Dreiner:2007vm}. Since 
mass reach is not a factor, one might suspect that the high intensity facilities used as $B$-factories, would 
be more sensitive, but this is also not the case \cite{Choudhury:1999tn}.  

In order to avoid over-closing the universe  \cite{Hooper:2002nq,Bottino:2011xv,Belanger:2013pna}, a 
10\,GeV or lighter neutralino must decay via R-parity violating operators \cite{Dreiner:1997uz}. If the 
R-parity violating couplings are not too small, in this case, the most promising method to search for a light 
neutralino, is via the production of mesons.  The rate for the latter is so high, that the subsequent rare 
decay of the meson to the light neutralino via (an) R-parity violating operator(s) can be searched for 
\cite{Dedes:2001zia,Dreiner:2002xg,Dreiner:2009er}. This is  analogous to the production of neutrinos 
via $\pi$ or $K$-mesons. 

For a neutralino in the mass range of about 0.5 - 5~GeV, the newly proposed SHiP experimental facility 
\cite{Anelli:2015pba} seems ideal. It will consist of a high intensity 400\,GeV proton beam from the CERN SPS 
incident on a fixed target. 63.8 m down beam line there will a detector for long-lived heavy neutral particles. 
Two of us (HKD, DS) have performed a preliminary analysis in \cite{Alekhin:2015byh}, for a small set of R-parity
violating operators: the decay $D^+\to\tilde\chi^0_1\,\ell^+_i$ via the operator $\lambda^\prime_{i21}L_iQ_2\bar 
D_1$, and the decays $\tilde\chi_1^0\to (K^0\nu;K^\mp\ell^\pm_i)$ via $\lambda^\prime_{i21,i21}$. It is the 
purpose of this paper to extend this to all possible production modes and decay channels, and to determine the 
search sensitivity of the SHiP experiment. As an example, in earlier work \cite{Dedes:2001zia}, the production of 
$B$-mesons at NuTeV was considered. The mesons could decay as $B_{d,s}^0\to \tilde\chi^0_1\,\nu$ or $B^\pm
\to\tilde\chi^0_1\,\ell_i^\pm$ via $\lambda^\prime_{i13}$. The neutralinos could decay via the $L_iQ_1\bar D_3$ 
operators, or purely leptonically via $L_iL_j\bar E_k$ operators. These scenarios can also be tested at 
SHiP, where a sufficiently large number of $B$-type mesons is expected and we will show the sensitivity reach 
to the $LQ\bar D$ operators here. 

This paper is organized as follows. In Sec.~\ref{sec:rpv}, we briefly review supersymmetry with broken R-parity.
We also discuss the bounds on the operators relevant to our analysis. Constraints on the neutralino mass and 
motivations for the possibility of a very light neutralino follow in Sec.~\ref{sec:light-neutralino}. In 
Sec.~\ref{sec:ProdShip} we discuss the relevant neutralino production and decay channels we expect to be 
most relevant for SHiP observations of R-parity violation (RPV). In Sec.~\ref{sec:rpvinteractions} we set up the 
effective field theory need to compute the relevant meson and neutralino decays and compute the latter in general fashion.
With these results at hand, we specialize to the SHiP set-up in Sec.~\ref{sec:ObsNeut}
and discuss under what circumstances a neutralino could be detected. In Sec.~\ref{sec:simulation} we explain the methodology of our numerical study. The discussion of the 
sensitivity of the SHiP experiment to the existence of light, but unstable, neutralinos in several benchmark scenarios is given in Sec.~\ref{sec:results}. In Sect.~\ref{sec:lhc-estimate}, we discuss an estimate for possible effects of our benchmark 
scenarios at the LHC. We conclude in Sect.~\ref{sec:conclusion}.

\section{R-Parity Violation}
\label{sec:rpv}
\subsection{Introduction}
The minimal supersymmetric extension of the SM requires the introduction of an additional Higgs doublet.
The minimal set of pure matter couplings are then encoded in the minimal supersymmetric SM (MSSM)
superpotential
\begin{eqnarray}
W_{\mathrm{MSSM}}&=&(h_E)_{ij} L_iH_d\bar E_j + (h_D)_{ij} Q_iH_d\bar D_j \nonumber \\
&& + (h_U)_{ij} Q_iH_u\bar U_j +\mu H_dH_u\,.
\end{eqnarray}
Here $h_{E,D,U}$ are dimensionless $3\times3$ coupling matrices. The Higgs mixing $\mu$ has mass dimension
one. This superpotential is equivalent to imposing the discrete $\mathbf{Z}_2$ symmetry R-parity 
\cite{Farrar:1978xj}, or the $\mathbf{Z}_6$ proton hexality \cite{Dreiner:2005rd}, and results in a stable proton.
Both are discrete gauge anomaly-free, ensuring stability under potential quantum gravity corrections 
\cite{Krauss:1988zc}. A model restricted to the superpotential $W_{\mathrm{MSSM}}$ is called R-parity conserving. The 
advantage is that the LSP, usually the lightest neutralino, is an automatic WIMP dark 
matter candidate  \cite{Goldberg:1983nd}. However, no such dark matter has been observed to-date, motivating
the search for other forms of supersymmetry.

Instead in R-parity violating models, there are further possible terms in the superpotential
\begin{eqnarray}
W_{\mathrm{RPV}}&=&W_{LV}+ W_{BV}\,,\\
W_{\mathrm{LV}}&=&\lambda_{ijk} L_iL_j\bar E_k + \lambda_{ijk}^\prime L_iQ_j\bar D_k +\kappa_i L_i H_u\,,\\
W_{\mathrm{BV}}&=&\lambda_{ijk}^{\prime\prime} \bar U_i \bar D_j \bar D_k\,.
\end{eqnarray}
Imposing the discrete $\mathbf{Z}_3$ symmetry baryon triality \cite{Ibanez:1991hv,Dreiner:2005rd}, the
lepton-number violating terms in $W_{\mathrm{LV}}$ remain. The proton remains stable, since baryon-number is 
conserved, however the neutralino LSP is unstable and is no longer a dark matter candidate. This can be 
solved by introducing the axion to solve the problem of CP-violation in QCD. The supersymmetric partner, 
the axino, is then automatically a good dark matter candidate 
\cite{Chun:1999cq,Hooper:2004qf,Dreiner:2014eda,Colucci:2015rsa} and light neutrino masses are also 
generated automatically \cite{Hall:1983id,Nardi:1996iy,Hirsch:2000ef,Dreiner:2006xw,Dreiner:2011ft}. We 
thus consider this a well-motivated  model to investigate. At a given energy scale the bi-linear terms $\kappa_i 
L_i H_u$ can be rotated away, even for complex $\kappa_i,\,\mu$ \cite{Hall:1983id,Nardi:1996iy,Dreiner:2003hw}. 
This leaves us with the tri-linear couplings $\lambda_{ijk}$ and $\lambda^\prime_{ijk}$. In this work we focus on 
the latter couplings as they lead to neutralino production via the decays of mesons.

\subsection{Bounds on R-parity Violating Couplings}
\label{sec:bounds}

The operators $L_iQ_j\bar D_k$ which we investigate here all violate lepton number; thus there are strict 
bounds on the coupling constants $\lambda^\prime_{ijk}$, see for example the reviews 
\cite{Barger:1989rk,Bhattacharyya:1997vv,Allanach:1999ic,Barbier:2004ez,Kao:2009fg}. We briefly 
summarize here the existing bounds on the specific operators which we investigate in our benchmark 
scenarios in Sec.\,\ref{sec:results}. For our results on the sensitivity reach of SHiP, we focus on the cases 
$\lam'_{1jk}$, involving final state electrons, and $\lam'_{31k}$, which produces tau leptons. However, experimentally final state muons should be testable
at least as well as electrons. We thus also present the corresponding bounds on the couplings $\lam'_{2jk}$, which
are typically weaker. We take most of the numbers from the most recent review \cite{Kao:2009fg}.
\begin{eqnarray}
\lam'_{112} &< 0.03\;\displaystyle\frac{m_{\tilde s_r}}{100\,\mathrm{GeV}},\qquad \lam'_{212} &< 0.06\,\frac{m_{\tilde s_R}}
{100\,\mathrm{GeV}} .\\[3mm]
\lam'_{121} &< 0.2\;\displaystyle\frac{m_{\tilde c_L}}{100\,\mathrm{GeV}},\qquad \lam'_{221} &< 0.1\,\frac{m_{\tilde d_R}}
{100\,\mathrm{GeV}}, \\[3mm]
\lam'_{122} &< 0.2\;\displaystyle\frac{m_{\tilde c_L}}{100\,\mathrm{GeV}},\qquad \lam'_{222} &< 0.1\,\frac{m_{\tilde s_R}}
{100\,\mathrm{GeV}},\\[3mm]
\lam'_{131} &< 0.03\;\displaystyle\frac{m_{\tilde t_L}}{100\,\mathrm{GeV}},\qquad \lam'_{231} &< 0.18\,\frac{m_{\tilde b_L}}
{100\,\mathrm{GeV}},\\[3mm]
\lam'_{312} &< 0.06\;\displaystyle\frac{m_{\tilde s_R}}{100\,\mathrm{GeV}},\qquad \phantom{\lam'_{231}}&\\[3mm] 
\lam'_{313} &< 0.06\;\displaystyle\frac{m_{\tilde t_L}}{100\,\mathrm{GeV}},\qquad \phantom{\lam'_{231}}& 
\end{eqnarray}
The bound on $\lam'_{231}$ is from \cite{Allanach:1999ic}, as there is no bound quoted in \cite{Kao:2009fg}. 
Assuming an experimental neutrino mass bound $m_\nu<1\,$eV results in a bound of approximately $\lam'_
{122,222}<0.01 \sqrt{\tilde m/100\,\mathrm{GeV}}$ \cite{Hall:1983id,Banks:1995by,Allanach:1999ic}. This is 
however model dependent, as we know there must be further, possibly off-diagonal, contributions to the 
neutrino mass matrix. 

\section{A Light Neutralino}
\label{sec:light-neutralino}

The best lower mass limit on the lightest supersymmetric particle (LSP) neutralino is from
LEP \cite{Agashe:2014kda}
\begin{equation}
m_{\tilde\chi^0_1} > 46\,\mathrm{GeV}.
\label{eq:lep-bound}
\end{equation}
This uses the LEP-II chargino search to restrict the range of $\mu$ and the SU(2) soft breaking gaugino 
mass parameter $M_2$ and then assumes the supersymmetric GUT relation 
\begin{equation}
M_1=\frac{5}{3} \tan^2\theta_W\,M_2\approx \frac{1}{2} M_2\,,
\label{eq:GUT-rel}
\end{equation}
where $M_1$ is the U(1)$_Y$ soft breaking gaugino mass. The LEP-II searches translate into $\displaystyle 
M_1 \gsim 50\,$GeV. Performing the Takagi diagonalization of the neutralino mass matrix \cite{Dreiner:2008tw}
and scanning the parameters $M_1,\,M_2,\,\mu,\,\tan\beta$ over the allowed ranges, results in the bound of 
Eq.~(\ref{eq:lep-bound}). 

If the GUT assumption is dropped and $M_1,\,M_2$ are independent parameters, setting the 
determinant of the neutralino mass matrix to zero results in the relation
\begin{equation}
M_1=\frac{M_2 M_Z^2\sin(2\beta)\sin^2\theta_{\mathrm{W}}}
{\mu M_2-M_Z^2\sin(2\beta)\cos^2\theta_{\mathrm{W}}}\,.
\end{equation}
For real parameters this equation can always be solved \cite{Dreiner:2009ic}. Thus for given values of $\mu,
\,M_2,$ and $\tan\beta$ there is always a mass-zero singular value \cite{Dreiner:2008tw}, and thus a 
massless neutralino state.
A neutralino lighter than $\mathcal{O}(10\,\mathrm{GeV})$ is dominantly bino and does not couple directly to 
the $Z^0$-boson. Thus the bounds on a light neutralino from the invisible $Z^0$-width are avoided 
\cite{Choudhury:1999tn}. In fact to our knowledge all laboratory bounds are avoided \cite{Dreiner:2009ic}.

The strictest mass bounds on a stable, light neutralino are astrophysical. Supernova  
\cite{Kachelriess:2000dz,Dreiner:2003wh} or white dwarf cooling \cite{Dreiner:2013tja}, give
a lower mass bound: $m_{\tilde\chi^0_1} \gsim250\,$MeV, for selectron masses around 320\,GeV. In this case, 
too few neutralinos are produced, due to Boltzmann suppression, as the supernova temperature is about 30\,MeV.
A \textit{massless} neutralino is allowed for selectron masses above about 1275\,GeV or below 320\,GeV. For 
$M_{\tilde e}>1275\,$GeV also too few neutralinos are produced. For $M_{\tilde e}<320\,$GeV the neutralinos 
are trapped in the supernova, similar to neutrinos and must be included in the full supernova simulation. Since 
this has not been done to-date, the supernova does not give a reliable bound in this region.

Cosmologically the Cowsik-McClelland bound \cite{Cowsik:1972gh} on a very light neutrino translates into the 
{\it upper} neutralino mass bound \cite{Dreiner:2009ic}
\begin{equation}
M_{\tilde\chi^0_1} < 0.7\,\mathrm{eV}\,.
\end{equation}
The neutralino in this case provides hot dark matter, but not enough to negatively affect structure formation. The 
observed dark matter density must then originate elsewhere, for example from the axino.

Requiring the lightest neutralino to provide the observed dark matter results in a {\it lower} mass bound, the 
Lee--Weinberg bound \cite{Lee:1977ua}. The proper bound is obtained by scanning over the allowed 
supersymmetric parameter space, while dropping the relation in Eq.~(\ref{eq:GUT-rel}). This is thus an 
on--going process \cite{Hooper:2002nq}. The most recent bound including the Higgs--discovery data and 
also constraints from stau searches gives \cite{Lindert:2011td,Belanger:2013pna,Calibbi:2013poa,
Calibbi:2014lga}
\begin{equation}
M_{\tilde\chi^0_1}>24\,\mathrm{GeV}\,.
\end{equation}
Therefore the mass range
\begin{equation}
0.7\,\mathrm{eV}<M_{\tilde\chi^0_1}<24\,\mathrm{GeV}\,,
\end{equation}
is excluded for a stable neutralino LSP as it gives too much dark matter. A neutralino LSP in this mass range 
is only allowed if it decays, \textit{i.e.} R-parity is violated.

\section{Light Neutralinos at SHiP}
\label{sec:ProdShip}

\subsection{The SHiP Setup}
\label{sec:ship}

We present some details of the SHiP setup that are relevant to our analysis. The SHiP proposal 
\cite{Anelli:2015pba} is not definitive yet. The plan is to employ the 400\,GeV proton beam at CERN in the 
fixed-target mode. This yields a center-of-mass energy of roughly \unit[27]{GeV}, sufficient to produce $D$ and $B$ mesons. Over the lifetime of the experiment a total of $2\cdot10^{20}$ protons on target are foreseen. Such a large event yield is expected to be achievable by \textit{e.g.} a hybrid target consisting of tungsten and titanium-zirconium doped molybdenum alloy. 

A major motivation for the SHiP experiment is to observe new, weakly-interacting particles with long lifetimes. Such particles could be produced via proton-target-collisions and propagate for finite distances of the order tens of meters before decaying back into Standard Model pairs. For that purpose, a decay volume is located 68.8\,m behind the target. It has a cylindrical shape with a total length of 
60 m, however with the first 5 m dedicated for background suppression vetoes. Furthermore, the decay region has an elliptic face front with semi-axes 5\,m and 2.5\,m. We sketch this setup in  
Fig.~\ref{fig:geometry}. A spectrometer and a calorimeter system positioned behind the decay volume can identify the visible final state particles that are potentially produced when a hidden particle decays.

\subsection{Production and Decay of Neutralinos via R-Parity Violation}

Since the goal is to investigate the light supersymmetric neutralino, one might consider their direct production. 
However, in Ref.~\cite{Dedes:2001zia} using the Monte Carlo program \texttt{HERWIG 6.2} \cite{Corcella:2000bw} 
the pair production of neutralinos in a proton fixed target experiment
\begin{equation}
p+p\to\tilde\chi^0_1\tilde\chi^0_1+X\,, \label{eq:pairwise}
\end{equation}
was shown to be three orders of magnitude too low at the NuTeV experiment, for squark masses of 100\,GeV. 
Since the light neutralino is almost pure bino, the dominant production mode is via $t$-channel squark exchange,
and the cross section goes as $1/m_{\tilde q}^4$. At SHiP it is planned to have 100 times more protons on target
than at NuTeV. However, the lower squark mass bound is now also about a factor of 10 stricter. One would thus
expect a further two orders of magnitude suppression, making this a hopeless endeavour also at SHiP.

\begin{figure*}
\includegraphics[width=0.3\textwidth]{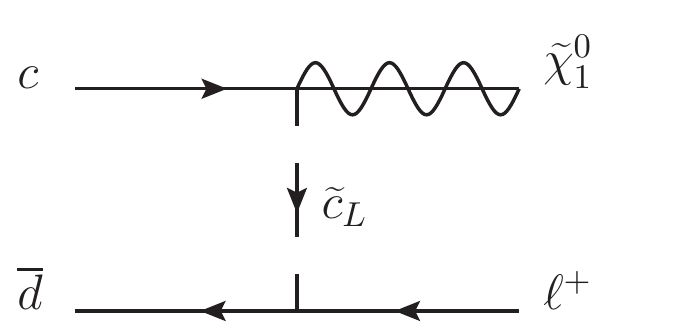} \qquad
\includegraphics[width=0.3\textwidth]{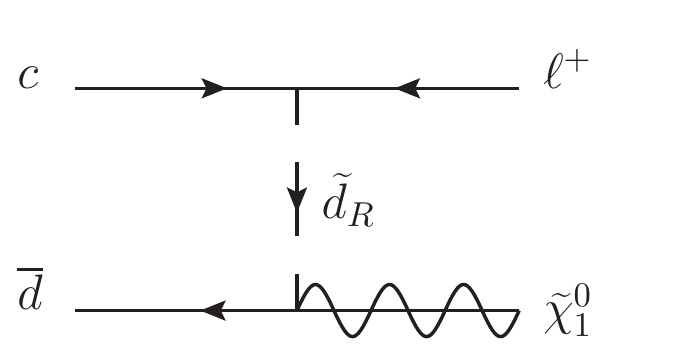} \qquad
\includegraphics[width=0.3\textwidth]{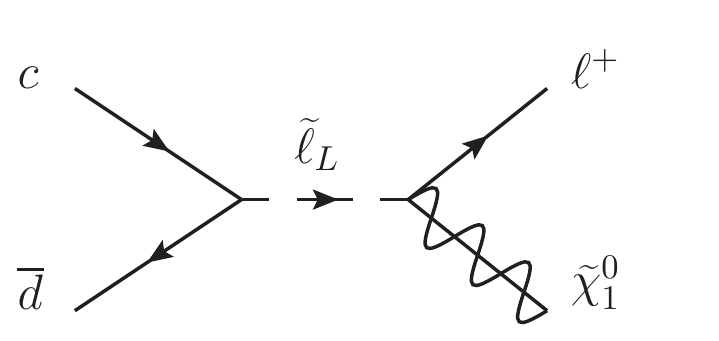}
\caption{Relevant Feynman Diagrams for $D^+ \rightarrow \neu + \ell^+$}
\label{fig:Ddecay}
\end{figure*}

As in Ref.~\cite{Dedes:2001zia}, we instead consider the production of mesons $M$, which 
can have a very large production cross section. On rare occasions these mesons can decay to the 
neutralino LSP and a neutral or charged lepton $l$ via
\begin{align}
p +  p &\to M + X \nonumber \\
 M &\to \tilde \chi^0_1 + l. \label{eq:singleneudecay}
\end{align}
At SHiP energies, with a 400\,GeV proton beam we expect (next to the production of light mesons containing 
up, down, or strange quarks) high production rates for charmed mesons, and somewhat lower rates for $B
$-mesons. As we discuss below, for example over the lifetime of SHiP about $4.8 \times 10^{16}$ $D^\pm
$-mesons are expected. Thus even very rare decays can be probed. Individual $L_iQ_a\bar D_b$ R-parity 
violating operators allow for leptonic decays of mesons. As an example the tree-level Feynman diagrams 
for the decay 
\begin{equation}
D^+\rightarrow\neu+ 
\ell^+_i,\,i=1,2\,.
\label{eq:Ddecay-LQD}
\end{equation}
are given in Fig.~\ref{fig:Ddecay}, for $a=2,\,b=1$. In this specific example the light neutralino can decay via 
the same R-parity violating operator: 
\begin{equation}
\tilde\chi^0_1\to (\nu \,K^0_{S/L};\, \bar\nu \,\bar K^0_{S/L})\,.
\label{eq:neut-decay-K}
\end{equation}
Both sets of decays are possible, as the neutralino is a Majorana fermion. For small values of the coupling 
$\lambda_{i21}^\prime$ and given that the neutralino must be lighter than the $D^+$ meson, the neutralino 
lifetime can be long enough to decay downstream in the SHiP detector.

\section{Effective Interactions: Lepton-Neutralino-Meson }
\label{sec:rpvinteractions}

In this section, we  discuss the R-parity violating effective interactions between a meson, a lepton, and a 
neutralino. These interactions are relevant for both the production and the decay of the neutralino and are 
necessary to determine the possible signatures at SHiP, as in 
Eqs.~(\ref{eq:Ddecay-LQD}),(\ref{eq:neut-decay-K}). We focus on the operators  $\lam_{iab}^\prime L_i
Q_a\bar D_b$ where $i$ denotes the leptonic generation index and $a$ and $b$ the quark generation indices.
The index $b$ is always associated with a down-like $SU(2)$ singlet quark, whereas the index $a$ can refer 
to either an up-like or down-like $SU(2)$ doublet quark. If $a$ 
is up-like then $i$ corresponds to a charged lepton, \textit{i.e.} electron, muon, or tau, whereas if $a$ is down, 
$i$ corresponds to a neutrino. 

\subsection{The Formalism}
\label{sec:rpvinteractions:formalism}
The interaction Lagrangian due to $\lam_{iab}^\prime L_iQ_a\bar D_b$
is given in terms of four-component fermions by
\begin{align}
 \mathcal{L} \supset&\, \lam_{iab}^{\prime} \Big[ (\overline{\nu^{\cal{C}}_i} 
 P_L d_a) {\widetilde d_{bR}}^* + (\overline{d_b} P_L \vi) {\widetilde d_{aL}} 
+ (\overline{d_b} P_L d_a) {\widetilde\nu}_{iL}  \Big] \nonumber \\
&- \lam_{iab}^{\prime} \Big[ (\overline{u_a^\cc} P_L \li) {\widetilde d_{b_R}}^* + (\overline{d_b} P_L u_a) 
 {\widetilde\ell_{iL} } + (\overline{d_b} P_L \li) {\widetilde u_{aL}} \Big]  \nonum \\
& + \text{h.c.}
 \label{eq:lqd} 
\end{align}
Here, $d_{a,b},\,u_a,\,\nu_i,\,\ell_i$ denote the down-like quark, up-like quark, neutrino, and 
charged lepton fields, respectively. The tilde denote the corresponding supersymmetric scalar partners.
The dominant contribution to the R-parity violating decay of a meson typically proceeds at tree-level via operators associated with the
Feynman diagrams as shown in Fig.~\ref{fig:Ddecay}. Thus we also need the standard supersymmetric
fermion-sfermion-neutralino vertices. We assume that the sfermion mixing is 
identical to the fermion mixing such that their contributions to the gauge couplings cancel. As we consider a 
dominantly bino LSP neutralino,  $\tilde\chi_1^0$, we only take chirality-conserving terms into account 
\begin{alignat}{3}
\mathcal{L} \supset & \;\;\;\;\,    g_{\tilde u_{aL} } (\overline{\neu} P_L u_a) {\widetilde u_{a_L}}^*  &&+ g_{\tilde d_{aL} } 
(\overline{\neu} P_L d_a) {\widetilde d_{a_L}}^*  \nonum \\
&+ g_{\tilde l_{iL} } (\overline{\neu} P_L \ell_i) {\widetilde\ell_{iL}}^* &&+  g_{\tilde \nu_{iL} } 
(\overline{\neu} P_L \nu_i) {\widetilde \nu_{i_L}}^* \nonum \\ 
&+ \gneudR^* (\overline{d} P_L \neu) {\widetilde d}_R &&+ \text{h.c.}
\end{alignat}
We assume that the sfermion masses are significantly larger than the momentum exchange of the process.
Thus the sfermions can be integrated out, resulting at tree-level in the low-energy effective 
four-fermion Lagrangian for  both the production and decay of the neutralino: 
\begin{eqnarray}
&&\!\!\!\!\!\!\!\!\!\!\!\!\!
\mathcal L \supset \lambda^\prime_{iab}\bigg[ \frac{g^*_{\widetilde d_{bR} }}{m^2_{\widetilde  d_{b_R}}}
(\bar d_b P_L \tilde \chi_1^0)(\overline{\nu^{\cal{C}}_i} P_L d_a)
\nonumber \\
&-&\!\!\frac{g^*_{\widetilde d_{bR} }}{m^2_{\widetilde  d_{b_R}}}(\bar d_b P_L \tilde \chi_1^0)(\overline{u^{\cal{C}}_a} P_L \ell_i)
+ \frac{g_{\widetilde d_{aL} }}{m^2_{\widetilde  d_{a_L}}}(\overline{ \tilde \chi^0_1} P_L d_a)(\overline{d_b} P_L \nu_i) 
\nonumber\\
&-&\!\! \frac{g_{\widetilde u_{aL} }}{m^2_{\widetilde  u_{a_L}}}(\overline{ \tilde \chi_1^0} P_L u_a)(\overline{d_b} P_L \ell_i) 
 + \frac{g_{\widetilde \nu_{iL} }}{m^2_{\widetilde  \nu_{i_L}}}(\overline{ \tilde \chi_1^0} P_L \nu_i)(\overline{d_b} P_L d_a) 
 \nonumber\\
&-&\!\! \frac{g_{\widetilde \ell_{iL} }}{m^2_{\widetilde  \ell_{i_L}}} (\overline{ \tilde \chi_1^0} P_L \ell_i)(\overline{d_b} P_L u_a) 
\bigg] + \mathrm{h.c.}
\label{eq:master-Lag}
\end{eqnarray}
We have omitted the terms involving pairs of neutralinos, which are most likely not 
relevant at SHiP, see \cite{Dreiner:2009er}. Similarly, we have dropped  interactions 
involving four SM fermions, see \cite{Barger:1989rk}.

For pure bino interactions, the coupling constants $g_X$ are family independent \cite{Drees:873465}
\begin{eqnarray}
g_{\tilde \ell_{iL}}&=& g_{\tilde \ell_{L}} = + \frac{g_2}{\sqrt{2}}  \tan \theta_W , \label{eq:gaugestart2} \\
g_{\tilde \nu_{iL}}&=& g_{\tilde \nu_{L}} = + \frac{g_2}{\sqrt{2}}    \tan \theta_W,  \\
g_{\tilde u_{aL}}&=& g_{\tilde u_{L}} = - \frac{g_2}{3 \sqrt{2}}  \tan \theta_W,  \\
g_{\tilde d_{aL}}&=& g_{\tilde d_{L}} = + \frac{5 g_2}{3 \sqrt{2}}   \tan \theta_W,  \\       
g_{\tilde d_{bR}}&=& g_{\tilde d_{R}} = - \frac{2 g_2}{3 \sqrt{2}}  \tan \theta_W, \label{eq:gaugeend2}
\end{eqnarray}
Here, $\theta_W$ denotes the electroweak mixing angle and $g_2$ the Standard Model SU(2) gauge coupling.

Using chiral Fierz identities (\textit{e.g.} \cite{chiralfierz}), one can rearrange the four-fermion interactions in 
Eq.~(\ref{eq:master-Lag}) such that each term factorizes in a neutralino-lepton current and a quark-bilinear:
\begin{align}
(\bar{\psi}_1 P_{\overset{L}{R}} \eta_2) (\bar{\eta}_1 P_{\overset{L}{R}} \psi_2) &= -\frac{1}{2} (\bar{\psi}_1 P_{\overset{L}{R}} \psi_2) (\bar{\eta}_1 P_{\overset{L}{R}} \eta_2) \nonumber \\
&- \frac{1}{4} (\bar{\psi}_1 \sigma^{\mu \nu} \psi_2) (\bar{\eta}_1 \sigma_{\mu \nu} \eta_2)  \nonumber \\
&\pm \frac{i}{8}  \epsilon^{\mu \nu \rho \sigma}(\bar{\psi}_1 \sigma_{\mu \nu} \psi_2) (\bar{\eta}_1 \sigma_{\rho \sigma} \eta_2), \label{eq:chiralfierz}
\end{align}
with $\sigma^{\mu \nu} \equiv i/2\ [\gamma^\mu, \gamma^\nu]$ and $\epsilon^{0123} = 1$. $\psi_{1,2},\,\eta_{1,2}$
denote four component fermions. Making use of these identities and applying $\overline{\psi^\cc} P_{L/R} \eta = 
\overline{\eta^\cc} P_{L/R} \psi$ in combination with the Majorana identity $\chi^\cc = \chi$ for the neutralino, 
Eq.~(\ref{eq:master-Lag}) can be written as the sum of the following four interactions:
\begin{align}
(\overline{\widetilde{\chi}^0} &P_L  \nu_i)  (\overline{d_b} P_L  d_a)\ \times \nonumber\\
&\underbrace{\lambda^\prime_{iab} \Big( \frac{ g_{\tilde \nu_{L} }}{m^2_{\tilde  \nu_{i_L}}}  -
 \frac{1}{2} \frac{g_{\tilde d_{L} }}{m^2_{\tilde  d_{a_L}}} - \frac{1}{2} 
\frac{g^*_{\tilde d_{R} }}{m^2_{\tilde  d_{b_R}}}\Big)}_{\equiv 
G^{S,\nu}_{iab}}, \label{eq:interactions1}\\[3mm]
(\overline{\widetilde{\chi}^0} &P_L  \ell_i) (\overline{d_b} P_L  u_a)\ \times \nonumber\\
&\underbrace{\lambda^\prime_{iab} \Big(\frac{1}{2} 
\frac{g_{\tilde u_{L} }}{m^2_{\tilde  u_{a_L}}} + \frac{1}{2} \frac{g^*_{\tilde d_{R} }}{m^2_{\tilde  d_{b_R}}} - \frac{ g_{\tilde \ell_{L} }}{m^2_{\tilde  \nu_{i_L}}}\Big)}_
{\equiv G^{S,\ell}_{iab}}, \label{eq:interactions2}\\[3mm]
(  \overline{\widetilde{\chi}^0} &\sigma^{\mu \nu}  \nu_i) (\overline{d_b} \sigma^{\rho \sigma}  d_a)\ \times \nonumber\\
&\underbrace{\lambda^\prime_{iab} \Big( \frac{g_{\tilde d_{L} }}{4 m^2_{\tilde  d_{a_L}}}  + \frac{g^*_{\tilde d_{R} }}
{4 m^2_{\tilde  d_{b_R}}} \Big)}_{\equiv G^{T,\nu}_{iab}} \Big( g_{\mu \rho} g_{\nu \sigma} - \frac{i \epsilon_{\mu \nu 
\rho \sigma}}{2}\Big), \label{eq:interactions3}
\end{align}
\begin{align}
(   \overline{\widetilde{\chi}^0} &\sigma^{\mu \nu}  \ell_i) (\overline{d_b} \sigma^{\rho \sigma}  u_a)\ \times \nonumber\\
&\underbrace{\lambda^\prime_{iab}\Big( \frac{g_{\tilde u_{L} }}{4 m^2_{\tilde  u_{a_L}}}  + \frac{g^*_{\tilde d_{R} }}
{4 m^2_{\tilde  d_{b_R}}} \Big)}_{\equiv G^{T,\ell}_{iab}} \Big( g_{\mu \rho} g_{\nu \sigma} - \frac{i \epsilon_{\mu \nu 
\rho \sigma}}{2}\Big), \label{eq:interactions4}
\end{align}
and their hermitean conjugates. 

For pseudoscalar mesons composed of anti-quarks $\bar q_1$ and quarks $q_2$, we can connect 
the quark bilinear vector currents with external meson fields by defining pseudoscalar meson decay constants 
$f_M$ 
\begin{equation}
\langle 0 | \bar q_1 \gamma^\mu \gamma^5 q_2 | M(p)\rangle \equiv i p^\mu f_M, \label{eq:vectormesonconstant1}
\end{equation}
where $|M(p)\rangle$ denotes a pseudoscalar meson $M$ with momentum $p$. The standard current-algebra 
approximation then predicts 
\begin{equation}
\langle 0 | \bar q_1 \gamma^5 q_2 | M(p_M)\rangle = i \frac{m_M^2}{m_{q_1} + m_{q_2}} f_M \equiv f^S_M, 
\label{eq:scalarmesonconstant}
\end{equation}
for the pseudoscalar currents in Eqs.\,\eqref{eq:interactions1} and \eqref{eq:interactions2}. Here, $m_M$, $m_{q_1}$ and $m_{q_2}$ are the masses of the meson $M$ and the quarks $q_1$, $q_2$, respectively.

The tensor structure in Eqs.\,\eqref{eq:interactions3},\,\eqref{eq:interactions4} does not lead to purely leptonic 
processes such as $M\rightarrow \widetilde{\chi}^0+l_i$ or $\widetilde{\chi}^0\rightarrow M + l_i$, because a 
pseudoscalar meson only has one relevant Lorentz-vector, its momentum. Thus the tensor interactions only 
contribute to higher multiplicity processes such as $M\rightarrow \widetilde{\chi}^0+l_i+M'$, where $M'$ denotes 
a lighter meson. These are phase space suppressed by two to three orders of magnitude, and we do not consider 
them here.

Vector mesons have two intrinsic Lorentz vectors, their momentum $p^\mu$ and polarization 
$\epsilon^\mu$. The decay constant of a vector meson $M^{*}$ with mass $m_{M^*}$, can be 
defined as
\begin{equation}
\langle 0 | \bar q_1 \gamma^\mu q_2 | M^{*}(p,\epsilon)\rangle \equiv f^V_{M^*} m_{M^*} \epsilon^\mu . 
\label{eq:vectormesonconstant2}
\end{equation}
Heavy-quark symmetry relates the vector and pseudoscalar constants for mesons containing a heavy quark $f^V_{M^*} 
\simeq f_M$ \cite{Isgur:1989vq}. We use this relation for the $B$ and $D$ mesons. For lighter mesons, such as $K^*$ and 
$\phi$, the relation is not accurate and instead we follow Ref.~\cite{Dreiner:2006gu}, where the vector decay constants 
are obtained from $M^*\rightarrow e^+ e^-$ decays. 

Similarly we can define the tensor meson constant
\begin{equation}
\langle 0 | \bar q_1 \sigma^{\mu\nu} q_2 | M^*(p,\epsilon)\rangle \equiv i f^T_{M^*} (p_M^\mu \epsilon^\nu-p_M^\nu \epsilon^\mu ).  
\label{eq:tensormesonconstant}
\end{equation}
For mesons containing a heavy quark ($c$ or $b$), heavy-quark symmetry \cite{Casalbuoni:1996pg} also relates the 
vector and tensor decay constants $f^T_{M^*} \simeq f^V_{M^*} \simeq f_M$. Because the tensor decay constants 
are not known in all cases, we also employ this relation for lighter mesons. The additional uncertainties entering via 
these simplifying assumptions hardly affect the SHiP sensitivity curves on $\lambda^\prime/m_{\widetilde{f}}^2$. 
These range over many orders of magnitude and thus an $\mathcal{O}(40\%)$ correction in a decay constant 
$f_{M^{(*)}}^{S,T,V}$ does not noticeably change the results presented in the figures in  Sec.~\ref{sec:results}.

We list the values of the pseudoscalar and vector decay constants we use in Table \ref{Tab:decay-constants} in
Sect.~\ref{sec:results}. In general, we find that neutralinos can interact both with pseudoscalar and vector 
mesons via different but related effective couplings. In the following analysis we therefore consider both meson 
types and also show how the inclusion of the latter affects the overall sensitivity. 

Potentially fine-tuned models with non-degenerate sfermion masses could lead to a complete 
cancellation of the individual contributions in Eqs.\,(\ref{eq:interactions1}),\,(\ref{eq:interactions2}). No sensitivity would 
be expected in such a scenario if only pseudoscalar mesons were considered in the analysis. However, for a nonzero 
RPV coupling the effective operators in Eqs.\,(\ref{eq:interactions1})--(\ref{eq:interactions4}) can not all vanish 
simultaneously. We hence safely use the simplifying assumption of completely mass 
degenerate sfermions.

\subsection{Possible Decay Modes}
\label{sec:decay}
From Eqs.\,(\ref{eq:interactions1})--(\ref{eq:interactions4}), a single $\lam^\prime_{iab}L_iQ_a \bar D_b$ operator leads 
to interactions with charged (pseudoscalar or vector) mesons $M^+_{ab}$ of flavour content $(u_a \overline{d_b}  )$, as 
well as neutral mesons $M^0_{ab}$ with quark composition $d_a \overline{d_b}$, and their respective charge conjugated 
equivalents. If $m_{\tilde\chi^0_1} < m_M-m_{l_i}$, the operator opens a decay channel of the meson into the neutralino 
plus lepton $l_i$. For example, $M^\pm \rightarrow \tilde\chi^0_1\ell^\pm$, or $M^0 \rightarrow \tilde\chi^0_1\nu,\;  \bar{M}
^0 \rightarrow \tilde\chi^0_1\bar{\nu}$. Such processes serve as the initial neutralino production mechanism here. In 
addition, for $m_{\tilde\chi^0_1} > m_M+m_{l_i}$ the neutralino can decay via $\tilde\chi^0_1 \rightarrow M^+ \ell^-, M^- 
\ell^+$ or $\tilde\chi^0_1\rightarrow M^0 \nu, \bar{M}^0 \bar\nu$. Such decays can potentially be observed in the 
SHiP detector.

From the structure of the operators in Eqs.(\ref{eq:interactions1})--(\ref{eq:interactions4}), the definition of the 
effective couplings and the meson structure constants in 
Eqs.\,(\ref{eq:scalarmesonconstant}),\,(\ref{eq:tensormesonconstant}), we obtain the following unpolarized decay 
widths\footnote{\label{footnote1}The neutralino decay width in Ref.~\cite{Alekhin:2015byh} contains an 
erroneous sign and misses a factor of 2, which is fixed here in Eq.~(\ref{eq:width3}).}:
\begin{widetext}
\begin{align}
\Gamma(M_{ab} \rightarrow \neu + l_i) &= \frac{\lambda^{\frac{1}{2}}(m_{M_{ab}}^2, m_{\neu}^2, m_{l_i}^2)}
{64 \pi m_{M_{ab}}^3} |G^{S,f}_{iab}|^2 (f^{S}_{M_{ab}})^2 (m_{M_{ab}}^2 - m_{\neu}^2 - m_{l_i}^2 ), \label{eq:width1}\\
\Gamma(M^{*}_{ab} \rightarrow \neu + l_i) &= \frac{\lambda^{\frac{1}{2}}(m_{M^*_{ab}}^2, m_{\neu}^2, m_{l_i}^2)}{3 
\pi m_{{M^*_{ab}}}^3} |G^{T,f}_{iab}|^2 (f^{T}_{M^*_{ab}})^2 \Big[m_{M^*_{ab}}^2 (m_{M^*_{ab}}^2 + m_{\neu}^2 + m_{l_i}^2) 
- 2 (m_{\neu}^2 - m_{l_i}^2)^2\Big], \\
\Gamma(\neu \rightarrow M_{ab} + l_i) &=  \frac{\lambda^{\frac{1}{2}}(m_{\neu}^2, m_{M_{ab}}^2, m_{l_i}^2)}{128 \pi 
m_{\neu}^3} |G^{S,f}_{iab}|^2 (f^{S}_{M_{ab}})^2 (m_{\neu}^2 + m_{l_i}^2 - m_{M_{ab}}^2), \label{eq:width3}\\
\Gamma(\neu \rightarrow M^*_{ab} + l_i) &=  \frac{\lambda^{\frac{1}{2}}( m_{\neu}^2, m_{M^*_{ab}}^2, m_{l_i}^2)}
{2 \pi m_{\neu}^3} |G^{T,f}_{iab}|^2  (f^{T}_{M^*_{ab}})^2 \Big[ 2 (m_{\neu}^2 - m_{l_i}^2)^2 - m_{M^*_{ab}}^2 (m_{M^*_{ab}}^2 
+ m_{\neu}^2 + m_{l_i}^2)\Big]. \label{eq:width4}
\end{align}
\end{widetext}
Here, $l_i$ either denotes $\ell_i^\pm$ or $\nu_i$, depending on whether  $M_{ab}$ is charged or neutral. 
The phase space function $\lambda^{\frac{1}{2}}(x,y,z) \equiv \sqrt{x^2+y^2+z^2-2xy-2xz-2yz}$. The coefficients $G$ are defined in 
Eqs.\,(\ref{eq:interactions1})--(\ref{eq:interactions4}). For each of the
above decays there exists a charge-conjugated process with identical decay width. Here we list the most important mesons $M_{ab}$ that participate in 
each interaction for given $a, b$
\allowdisplaybreaks
 \begin{eqnarray}
 \lam^\prime_{i11} &\to& \left\{\begin{array}{l} (u\bar d) = (\pi^+,\rho^{+})\\[2mm]
 (d\bar d) = (\pi^0,\eta,\eta',\rho,\omega)\,,
 \end{array}
  \right. \\[2mm] \label{eq:lambdatomesonsstart}
 \lam^\prime_{i12}&\to& \left\{\begin{array}{l}  (u\bar s) = (K^+,\,K^{*+})\\[2mm]
  (d\bar s) = (K^0_L, K^0_S,K^{*0})\,,
 \end{array}
  \right. \\[2mm]
 \lam^\prime_{i13} &\to& \left\{\begin{array}{l} (u\bar b) =(B^+,\,B^{*+}) \\[2mm]
  (d\bar b) =  (B^0,\,B^{*0})\,,
 \end{array}\right. 
 \\[2mm]
 \lam^\prime_{i21} &\to& \left\{\begin{array}{l}  (c\bar d) = (D^+,\,D^{*+})\\[2mm]
  (s\bar d) = (K^0_L, K^0_S,K^{*0})\,,
 \end{array}
  \right. \\[2mm] 
  \quad\lam^\prime_{i22}&\to& \left\{\begin{array}{l}(c\bar s)= D_s^+,\,D_s^{*+}\\[2mm]
 (s\bar s)=\eta,\,\eta',\,\phi
 \end{array}
  \right. \\[2mm]
 \quad\lam^\prime_{i23}&\to& \left\{\begin{array}{l}(c\bar b)=B_c^+,\,B_c^{*+},\,
 \\[2mm]
 (s\bar b)= B_s^0,\,B_s^{*0}
 \end{array}\right. 
 \\[2mm]
 \quad\lam^\prime_{i31}&\to&  \phantom{\{ \ \ } (b\bar d)= B^0,\,B^{*0}  \\[2mm]
 \quad\lam^\prime_{i32}&\to&  \phantom{\{ \ \ }(s\bar b)= B_s^0,\,B_s^{*0} \\[2mm]
 \quad\lam^\prime_{i33}&\to&  \phantom{\{ \ \ } (b\bar b)=\eta_b,\, \Upsilon \label{eq:lambdatomesonsend}
 \end{eqnarray}
 
For light neutral pseudoscalar mesons, mass and flavour eigenstates do not coincide. For our studies this is only 
relevant for the $\bar K^0_{L,S}$, $\eta$, and $\eta^\prime$ mesons (we take $\phi$ to be a pure ($s \bar s$) state).
For the former, we neglect any CP-violation and define the mass eigenstates $|K_{L/S}\rangle \equiv (|K_0\rangle\pm|\bar 
K_0\rangle)/\sqrt 2$, where $|K_0\rangle$ and $|\bar K_0\rangle$ are flavor eigenstates $(d \bar s)$ and $(s \bar d)$, 
respectively. We can then read off the decay constants from
\begin{align}
\hspace{-3.5pt}\langle 0 | \bar s \gamma^\mu \gamma^5 d | K^0_{L}(p)\rangle=& + \langle 0 | \bar d \gamma^\mu \gamma^5 s
 | K^0_{L}(p)\rangle =   \frac{i p^\mu f_K}{\sqrt{2}}, \\
\hspace{-3.5pt}\langle 0 | \bar s \gamma^\mu \gamma^5 d | K^0_{S}(p)\rangle=& - \langle 0 | \bar d \gamma^\mu \gamma^5 s
 | K^0_{S}(p)\rangle =  \frac{i p^\mu f_K}{\sqrt{2}},
\end{align}
where $f_K$ is the decay constant of the charged kaon as defined in Eq.~\eqref{eq:vectormesonconstant1}.

For $\eta$ and $\eta^\prime$ we consider mixing between the $\eta^0$ and $\eta^8$ flavor states. We are only interested 
in the $(\bar s s)$ content of these mesons. We follow Refs.~\cite{Feldmann:1998vh,Dreiner:2009er}  and define
\begin{equation}
\langle 0 | \bar s \gamma^\mu \gamma^5 s | \{\eta,\,\eta'\}(p)\rangle= i p^\mu f^{\bar s s}_{\{\eta,\,\eta'\}}\,,
\end{equation}
and give numerical values in Table \ref{Tab:decay-constants}.

For the special cases $\lam^\prime_{ijj}$ the radiative neutralino decay is possible 
\cite{Hall:1983id,Dawson:1985vr}
\begin{equation}
\tilde\chi^0_1\to \gamma + (\nu_i,\,\bar\nu_i)\,.
\end{equation}
This is necessarily relevant for very light neutralinos, below the pion mass. We do not know how well this 
would be visible at SHiP, and do not consider it further here.

\begin{table}
\begin{tabular}{l@{\hspace{1cm}}r@{\hspace{1cm}}l}
\toprule
Decay    & Value & Ref. \\
constant &       &      \\
\hline
$f^{\bar s s}_\eta$ & -142 MeV & \cite{Feldmann:1998vh,Dreiner:2009er} \\
$f^{\bar s s}_{\eta^\prime}$ & 38 MeV & \cite{Feldmann:1998vh,Dreiner:2009er} \\
$f^V_\phi$ & 230 MeV & \cite{Dreiner:2006gu} \\
$f_K$& 156 MeV & \cite{Agashe:2014kda}\\
$f^V_{K^*}$ & 230 MeV & \cite{Dreiner:2006gu} \\
$f_D,\,f^V_{D^\star}$ & 205 MeV & \cite{Agashe:2014kda} \\
$f_{D_s},\,f^V_{D_s^\star}$ & 259 MeV & \cite{Agashe:2014kda} \\
$f_B$ & 191 MeV & \cite{Agashe:2014kda} \\
$f_{B_s}$ & 228 MeV & \cite{Aoki:2013ldr} \\
\botrule
\end{tabular}
\caption{Values of the pseudoscalar and vector decay constants that are used in the various benchmark scenarios. 
Definitions of the constants are given in Eqs.~(\ref{eq:vectormesonconstant1}) and (\ref{eq:vectormesonconstant2}). 
Tensor decay constants are chosen to be equal to the pseudoscalar decay constants. }
\label{Tab:decay-constants}
\end{table}

\section{Observable Neutralinos}\label{sec:ObsNeut}

With the predicted decay widths at hand we investigate how and under which circumstances R-parity violation can 
be observed at the SHiP experiment. Each $\lambda^\prime$ coupling causes at least one type of meson to decay into a 
neutralino and another charged or neutral lepton, provided it is  kinematically allowed. This process serves as the 
initial neutralino production mechanism at SHiP in our analysis. Given the number $N_M$ of mesons $M$ 
produced at SHiP and the lifetime $\tau_M$, the expected number of initially produced neutralinos is given by 
\begin{align}
N_{\chi}^{\text{prod.}} = \sum_M N_M \cdot \Gamma(M \rightarrow \neu + l) \cdot\tau_M. \label{eq:chiprod}
\end{align}
As apparent from the previous section, for each operator there are both pseudoscalar and vector 
mesons which can produce neutralinos. However, the lifetimes of a pseudoscalar and the corresponding 
vector meson of the same quark composition differ by many orders of magnitude. As an example, for 
the lightest charged charm meson, $D^\pm$, and its vector resonance partner, $D^{*\pm}$, one finds $\tau_{D^{*
\pm}} / \tau_{D^{\pm}} \approx 8 \times 10^{-9}$. Similar ratios appear for kaons and even though the lifetime of 
vector B-mesons is yet unknown there is no reason to expect largely different behaviour. For the RPV decay widths, 
however, one finds that $\Gamma(D^{*\pm} \rightarrow \neu \ell^\pm) / \Gamma(D^{\pm} \rightarrow \neu \ell^\pm)$ 
depends mainly on the ratio of masses and of the effective operator couplings $G_T/G_S$. Thus, it is hardly larger 
than 2 orders of magnitude unless one chooses a very peculiar setup of fine-tuned parameters. As the expected 
number of initial mesons, $N_{M}$ and $N_{M^*}$, will also be of roughly the same order, we conclude that if both $M
\to \neu f$ and $M^* \to \neu f$ are kinematically allowed, the contribution from vector meson decays is completely 
negligible. 

In the small mass range $m_{M} < m_{\neu} < m_{M^*}$ it might only be the vector mesons that can produce 
neutralinos in the first place, but by the above arguments and from the results below we expect the neutralino 
event rates to be far too small to be observable. We therefore ignore any neutralino production 
via vector meson decays in the following study.

For the neutralinos to be observable, a sufficiently high fraction must decay within the decay 
chamber of the SHiP experiment. By summing the widths $\Gamma(\neu \rightarrow M f)$ of all allowed 
channels for a given operator, we can derive the proper lifetime of a neutralino given the parameters of 
the RPV supersymmetric model. Given the kinematical distributions of neutralinos produced via meson 
decay and knowing the geometry of the decay chamber, we can find the average probability $\langle P [\tilde\chi^0_1
\text{ in d.r.}] \rangle$ of a neutralino decaying inside the detectable region. This is explained in more detail below.

A neutralino decaying inside the decay chamber is a necessary, but not a sufficient condition, as the final state 
particles have to be observed and traced back to an invisibly decaying new particle. For a charged final state, 
\textit{e.g.} $K^+ e^-$, one can measure the trajectory of both particles, measure their momenta and presumably 
identify the neutralino decay vertex. For a neutral final state, \textit{e.g.} $K_L \nu$, one loses both the tracking 
information and the momentum of the second particle. We expect that these are hard to be linked to the decay of 
a neutralino. Thus we only count neutralinos that decay into charged final state particles\footnote{We note that 
this assumption has not been made in \cite{Alekhin:2015byh} and as such, our results for the same scenario 
slightly differ.}\!\!\!. The final number of observed neutralinos is then
\begin{align}
\hspace{-0.35cm}N_{\tilde\chi^0_1}^{\text{obs.}} = N_{\tilde\chi^0_1}^{\text{prod.}} \cdot \langle P [\tilde\chi^0_1 
\text{ in d.r.}] \rangle \cdot \text{BR}(\tilde\chi^0_1 
\to \text{charged})\,. \hspace{-0.25cm} \label{eq:chidec}
\end{align}
With the above considerations, we thus demand the following for observable neutralino decays via $LQD$ 
operators at SHiP:
\begin{enumerate}
\item A pseudoscalar meson $M$ with $m_M > m_{\neu}$ must have a non-vanishing decay rate into neutralinos.
\item The neutralino must have a non-vanishing decay rate into another charged meson $M^{\prime \pm}$ with 
$m_{M^{\prime \pm}} < m_{\tilde\chi_1^0} < m_M$. 
\end{enumerate}
With these this conditions, it is practically impossible for SHiP to observe R-Parity violation if only one $\lambda^
\prime_{iab}$ coupling is nonzero. Eqs.~(\ref{eq:lambdatomesonsstart})--(\ref{eq:lambdatomesonsend}) show which 
operator leads to which sets of mesons that  can decay into the neutralino or the neutralino can decay into. The 
only operator related to both a pseudoscalar meson $M$ and a charged meson $M^{\prime \pm}$ with  $m_M > 
m_{M^{\prime \pm}}$ is $\lambda^\prime_{112}$, which might be observable via the chain 
\begin{equation}
K^0_{L/S} \rightarrow \tilde\chi^0_1 \nu,\qquad  \tilde\chi^0_1 \rightarrow K^\pm \ell^\mp\,.
\end{equation}
However, as $|m_{K^\pm} - m_{K^0_{L/S}}| \approx 4$ MeV, the testable range of neutralino masses is extremely 
limited and the expected energies of the final state particles are so small that the decays would be very challenging to observe.

Thus we require two different operators $\lambda^\prime_{iab}, \lambda^\prime_{jcd}\not=0$, with 
$iab$ and $jcd$ such that the decays fulfill the above requirements. This necessary extension leads to a plethora 
of possible combinations. In Sect.~\ref{sec:results} we restrict ourselves to an interesting subset of benchmark scenarios.

\section{Simulation of RPV Scenarios}
\label{sec:simulation}
Eqs.~(\ref{eq:chiprod}),\,(\ref{eq:chidec}) tell us how to estimate the number of observable neutralino decays 
$N_{\tilde\chi^0_1}^\text{obs}$ for any given operator combination and parameter values. The total widths 
and branching ratios into charged final states can be calculated from the general width formulae, 
Eqs.~(\ref{eq:width1})-(\ref{eq:width4}). We next describe the numerical tools we use to estimate $N_M$ and 
$\langle P [\tilde\chi^0_1 \text{ in d.r.}] \rangle$.
 
To get a reliable estimate on the kinematics of the initially produced mesons at SHiP, as well as the resulting 
neutralinos after their decay, we use \verb@Pythia 8.175@ \cite{Sjostrand:2006za, Sjostrand:2007gs}. In 
scenarios with initial charm (bottom) mesons we use the \texttt{HardQCD:hardccbar} (\texttt{HardQCD:hardbbbar}) 
matrix element calculator within \texttt{Pythia}, which includes the partonic processes $q \bar{q}, g g \to c \bar{c}\,
(b \bar{b})$ and select the specific meson type for each benchmark scenario. According to \cite{Bonivento:2013jag},
the number of $c \bar{c}$ events after 5 years of operation is expected to be $N_{c\bar{c}} = 9 \times 10^{16}$. By 
simulating 1M events of type $c \bar{c}$, we can use \texttt{Pythia} to find the average number of produced 
charm mesons per $c\bar{c}$ event, \textit{i.e.} 
\begin{equation}
n^{c\bar c}_{D} \equiv N_{D} / N_{c\bar{c}}, \;\;\mathrm{with} \;\;D \in \{D^\pm, D_s\}\,.
\end{equation} 
The analogous simulation of $b \bar{b}$ 
events gives the respective number for bottom mesons: 
\begin{equation}
n^{b\bar b}_{B} \equiv N_{B} / N_{b\bar{b}}, \;\;\mathrm{with}\;\;B \in \{B^0, B^\pm, B^0_s\}\,.
\end{equation}
The total number of expected $b \bar{b}$ events is taken by 
scaling the known number for $c \bar{c}$ events by the ratio of total cross sections determined by \texttt{Pythia}, 
\textit{i.e.} $N_{b\bar{b}} = N_{c\bar{c}} \times \sigma_{b\bar{b}} / \sigma_{c\bar{c}}$. We therefore combine
\begin{align}
N_M = N_{c\bar{c}} \cdot  \left\{
   \begin{array}{ll} n^{c\bar c}_M   &\text{ for charm mesons} \\[2mm]
     n^{b\bar b}_M \cdot \sigma_{b\bar{b}} / \sigma_{c\bar{c}}  &\text{ for bottom mesons}
    \end{array}
\right.  \label{eq:prodmes}
\end{align}
and list the numerical values in Table \ref{tbl:prodnumerics}.

\begin{table}
\begin{tabular}{lcl}
\toprule
$N_{c\bar{c}}$ & \quad & $9 \times 10^{16}$ \\[0.6mm]
$\sigma_{b\bar{b}} / \sigma_{c\bar{c}}$ & & $2.1 \times 10^{-4}$ \\ [0.9mm]
\hline
$ n^{c\bar c }_{D^\pm} $ & & 0.53 \\ [0.8mm]
$ n^{c\bar c }_{D_s^\pm}$ & & 0.074 \\ [1.1mm]
\hline
$ n^{b\bar b }_{B^\pm}$ & & 0.83 \\ [0.8mm]
$ n^{b\bar b }_{B^0}$ & & 0.80 \\[0.8mm]
$ n^{b\bar b }_{B_s^0}$ & & 0.14 \\
\botrule
\end{tabular}
\caption{Numerical values used to estimate the number $N_M$ in Eq.~(\ref{eq:prodmes}). Except for $N_{\bar{c} c}$, 
which is taken from  \cite{Bonivento:2013jag}, all numbers are evaluated by simulating 1M events of each 
\texttt{HardQCD} type in \texttt{Pythia}. }
\label{tbl:prodnumerics}
\end{table}

\begin{figure}
\vspace{-0.5cm}
\def\svgwidth{250pt} 
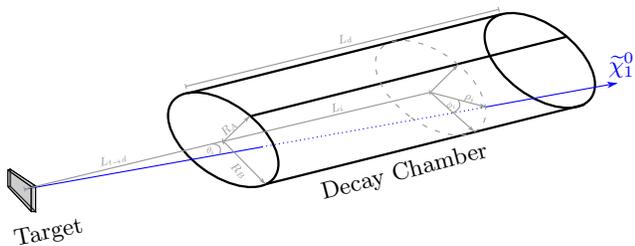
\vspace{-0.5cm}
\caption{Schematic overview of the SHiP detector geometry and definition of distances and angles used in text.}
\label{fig:geometry}
\end{figure}

For each benchmark scenario, we simulate $20,\!000$ events of the correct \texttt{HardQCD} type and --- to 
increase statistics --- set the branching ratio BR$(M \to \tilde\chi^0_1 f)$ to 100\,\%. We then scale our 
results accordingly. As the decaying mesons are scalar particles, the momentum of the neutralino is chosen to 
be uniformly distributed in the rest frame of the decaying meson.  We then sum over all produced neutralinos and 
determine the average probability, \textit{i.e.} for all possible neutralino momenta, that an arbitrary neutralino decays 
within the SHiP detector. Given the four-vector of the neutralino $(\tilde\chi^0_1)_i$ in spherical coordinates as
$(E_i, p^z_i, \theta_i, \phi_i)$ and the distances and angles as defined in Fig.~\ref{fig:geometry}, the average 
probability for $N^{\text{MC}}_{\tilde\chi^0_{1}}$ neutralinos in a given sample generated by a Monte Carlo program 
is evaluated as
\begin{eqnarray}
\langle P [\tilde\chi^0_1 \text{ in d.r.}] \rangle &=& \frac{1}{N^{\text{MC}}_{\tilde\chi^0_{1}}} \sum_{i = 1}^{N^{\text{MC}}_{\tilde\chi^0_{1}}} P[(\tilde\chi^0_{1})_i \text{ in d.r.}], 
\\[2mm]
P[(\tilde\chi^0_{1})_i \text{ in d.r.}] &= &e^{-L_{t \to d}/\lambda^z_i} \cdot \Big(1 - e^{-L_i/\lambda^z_i}\Big). \label{eq:observechi}
\end{eqnarray}
The mean decay length $\lambda^z_i$ of the neutralino in the lab frame is given by
\begin{eqnarray}
\lambda^z_i &=& \beta^z_i \gamma_i / \Gamma_{\mathrm{tot}}(\tilde\chi^0_1),  \label{eq:probdefinitions1}\\[2mm]
\beta^z_i &= & p^z_i / E_i, \label{eq:probdefinitions2} \\[2mm]
\gamma_i &= &E_i / m_{\tilde\chi^0_1}. \label{eq:probdefinitions3}
\end{eqnarray}
$\beta_i^z$ is the $z$-component of the relativistic velocity of $(\tilde\chi^0_1)_i$, $\gamma_i$ the corresponding 
Lorentz boost factor. $\Gamma_{\mathrm{tot}}(\tilde\chi^0_1)$ is the total decay width of the neutralino LSP, which 
only depends on the model parameters and not on the kinematics of an individual candidate $(\tilde\chi^0_1)_i$.
$L_i$ denotes the distance in $z$-direction a neutralino can travel inside the decay chamber before leaving it in radial direction.  
It can be determined as, see  also Fig.~\ref{fig:geometry},
\begin{eqnarray}
 L_i &= &\left\{ 
\begin{array}{ll}
   0 & \text{if $\rho_i \cot \theta_i < L_{t \to d}$}, \\[1.5mm]
   L_d & \text{if $\rho_i  \cot \theta_i > L_{t \to d} + L_d$}, \\[1.5mm]
   \rho_i \cot \theta_i - L_{t \to d} & \text{else}, 
\end{array}\right.  \\[2mm]
\rho_i & =& R_A R_B / \sqrt{(R_B \cos \phi_i)^2 + (R_A \sin \phi_i)^2}.
\end{eqnarray}
$\rho_i$ is the radius of the ellipse in the direction $\phi_i$. $\theta_i$ is the angle between the flight direction of 
the neutralino and the central axis of the detector; the polar angle, $\phi_i$, is the azimuthal angle of the 
neutralino momentum 3-vector. $R_A$ denotes the semi-minor axis, and $R_B$ the semi-major axis of the
elliptical face of the detector. $L_{t\to d}$ denotes the distance from the target to the front of the detector. $L_d$ 
denotes the length of the detector along the central axis. As explained in Sec.~\ref{sec:ship}, we use the 
numerical values $L_{t \to d} = 68.8$ m, $L_d = 55$ m, $R_A = 2.5$ m and $R_B = 5$ m.

\section{Results for various Benchmark Scenarios}
\label{sec:results}
In Eqs.~(\ref{eq:lambdatomesonsstart})-(\ref{eq:lambdatomesonsend}) we have listed the twenty-seven 
operators $\lambda_{iab}^\prime$ together with the corresponding mesons they couple to. For a fixed lepton 
flavor there are $36$ possible combinations for production and decay of the neutralinos, if we assume
distinct operators. The number of possibilities exceeds 100 if one in addition tests all possible values for the lepton flavor indices. It is clear that we can not investigate all of these cases in detail. In order to 
analyze the sensitivity at SHiP, we have thus focussed on a subset which we propose as, hopefully
representative, benchmark scenarios. 

\begin{figure*}
\begin{center}
\includegraphics[trim={0 -0.5cm 0  0},clip,height=0.25\textheight]{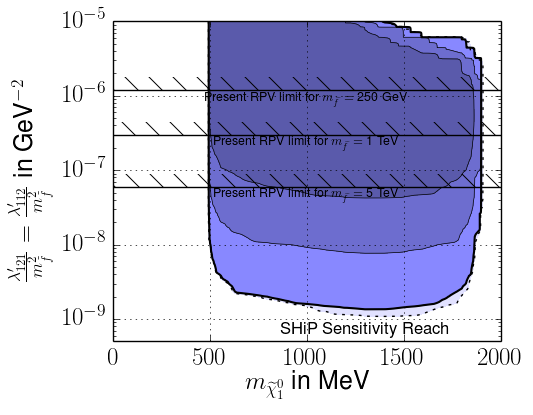} \qquad \qquad
\includegraphics[trim={0 0 0 0},clip,height=0.25\textheight]{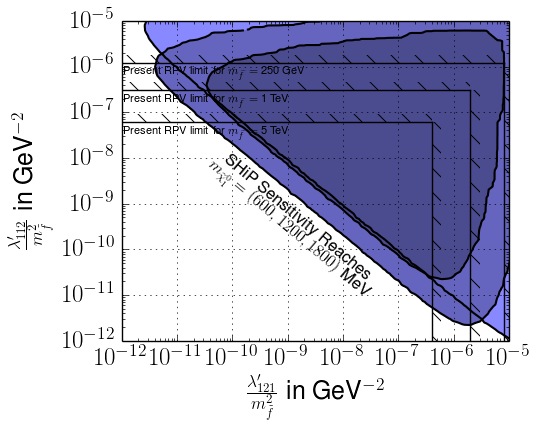} \\
\end{center}
\vspace{-1.25cm}
a) \hspace{0.45\textwidth} b) \hspace{0.4\textwidth} \hfill \\
\caption{SHiP sensitivity curves for Benchmark Scenario 1. In a), the two couplings are set
equal. The maximum sensitivity reach, corresponding to $\geq3$ events, is shown in bright blue, with a solid 
curve edge. The blue area corresponds to $\geq3\cdot10^3$ events and dark blue to $\geq3\cdot10^6$ events.
The dashed curve extending just below the light blue region denotes the extended sensitivity if the neutral
mesons from neutralino decays are also visible. The horizontal hashed lines correspond to the existing
limits on the RPV couplings, for three different sfermion masses: 250\,GeV, 1\,TeV, 5\,TeV. In b),
the sensitivity reach is shown as a function of the two independent RPV couplings for three fixed neutralino masses:
600\,MeV (bright blue), 1200\,MeV (blue), 1800\,MeV (dark blue). For the $x$-axis we always choose the coupling 
responsible for the production of the neutralinos, here $\lam^\prime_{121}/m^2_{\tilde{f}}$. For the $y$-axis we 
choose the coupling responsible for the decay of the neutralinos, here $\lam^\prime_{112}/m^2_{\tilde{f}}$. The 
existing bounds on the RPV couplings are again shown as solid hashed lines for 3 different sfermion masses: 
250\,GeV, 1\,TeV, 5\,TeV.}
\label{fig:121-112}
\end{figure*}

In choosing the benchmark scenarios, we took the following points into consideration. For the sensitivity, to
first order, it does not matter if we consider electrons or muons. We thus restrict ourselves to 
electrons.\footnote{Although the bounds on $\lam'_{2jk}$ are typically weaker than on $\lam'_{1jk}$.} We have 
one benchmark with final state taus, as their considerably larger mass affects the accessible decay 
phase space and the respective total widths.

The meson production rates can differ substantially. Thus we consider various scenarios where the neutralinos are
produced via neutral or charged $D$- or $B$-mesons. To estimate the production rates we use \texttt{Pythia}, as 
discussed above. We do not consider the production of neutralinos via $\lam_{i11}^\prime, \lam_{i12}^\prime$ as the 
production of the corresponding light mesons are not well simulated in forward direction with \texttt{Pythia}. We 
postpone this to future work. For kinematic reasons, we restrict the final state mesons in the neutralino decays to 
$K$ and $D$ mesons. 
We also do not consider decays into pions and the associated vector resonances, as we expect 
sizable deviations from our approximations. However, from the results of the benchmark scenarios discussed below, 
an estimate for pion final states can be derived easily, by letting the neutralino mass range down to the pion mass of 
about 135 MeV, instead of the kaon mass. The pion and kaon decay constants are related by $SU(3)$ flavor symmetry. Of course, a neutralino decay into pions requires turning on the coupling $\lambda^\prime_{i11}$ where $i=1,2$. Note that in this case it does matter if 
$i=2$ as now $m_\mu \simeq m_\pi$.

To be precise we consider the following cases, which we propose as benchmarks.

\subsection{Benchmark Scenario 1}
\begin{table}[thb]
\begin{tabular}{r||l}
\hline
$\lambda^\prime_P$ for production & $\lambda^\prime_{121}$ \\
$\lambda^\prime_D$ for decay & $\lambda^\prime_{112}$ \\
produced meson(s) & $D^\pm$ \\
visible final state(s) & $K^{(*)\pm} e^\mp$ \\
invisible final state(s) via $\lambda^\prime_P$ & $(K^0_L,K^0_S, K^*) + (\nu, \bar{\nu})$ \\
invisible final state(s) via $\lambda^\prime_D$ & $(K^0_L,K^0_S, K^*) + (\nu, \bar{\nu})$ \\
\hline
\end{tabular}
\caption{Features of Benchmark Scenario 1.}
\label{tab:sec-1}
\end{table}


We begin with scenarios where the neutralino is produced via the RPV decay of a $D$ meson and subsequently 
decays into a kaon plus lepton. This scenario has already been studied in some detail (by two of us, HD, DS) in 
Ref.\,\cite{Alekhin:2015byh}. We turn on two RPV couplings $\lambda_{121}^\prime$ and $\lambda_{112}^\prime$.
Neutralino production then occurs via the decay $D^{\pm}\to  \tilde\chi^0_1 + e^{\pm}$ which is proportional to 
$(\lambda_{121}^{\prime})^2$. The same coupling also leads to neutralino decay via the process $ \tilde\chi^0_1 
\rightarrow (K^0_L, K^0_S,K^{*0}) + \nu$ which contains no charged particles in the final state and will therefore be 
difficult to observe. However,  these decays do impact the neutralino lifetime. The relevant information is summarized in 
Table\,\ref{tab:sec-1}.

Because we have turned on the coupling $\lambda^\prime_{112}$ the neutralino can furthermore decay via 
$ \tilde\chi^0_1 \rightarrow (K^{\pm}, K^{*\pm}) + e^{\mp}$, which is possible to detect at SHiP. This coupling also leads 
to the same invisible decay to neutral kaons and neutrinos as $\lambda^\prime_{121}$. The invisible decays are 
important to include in the computation. 

We now present our results. The expected number of events depends on three independent parameters: $\lambda^
\prime_{121}/m_{\tilde{f}}^2$, $\lambda^\prime_{112}/m_{\tilde{f}}^2$, and the neutralino mass $m_{\tilde \chi^0_1}$. 
We find it convenient to present our results in two different ways. At first, we assume the RPV coupling constants to 
be equal, $\lambda^\prime_{121} = \lambda^\prime_{112}\equiv \lambda^\prime$. In Fig.~\ref{fig:121-112}a) we 
show the number of expected visible neutralino decays in the SHiP detector as event rate iso-curves which 
are functions of $\lambda^\prime/m_{\tilde f}^2$ and $m_{\tilde \chi^0_1}$. The bright blue area bounded by a thick 
solid line shows the expected maximum sensitivity curve for the SHiP experiment. This area in parameter space gives 
rise to $\geq 3$ neutralino decays within the detector into charged final state particles for $2\cdot10^{20}$ 
protons-on-target. The corresponding expected meson production rates are listed in Sec.~\ref{sec:simulation}. To show 
how the event rate increases with $\lambda^\prime/m_{\tilde{f}}^2$ and $m_{\tilde \chi^0_1}$, we also show areas for 
$\geq 3\cdot10^{3}$ (blue) and $\geq 3\cdot10^6$ (dark blue) observable decays. The horizontal hashed lines depict 
existing bounds on the couplings for various values of the sfermion mass $m_{\tilde{f}}$, \textit{cf.} Sect.~\ref{sec:bounds}.

\begin{figure*}
\begin{center}
\includegraphics[trim={0 -0.5cm 0  0},clip,height=0.25\textheight]{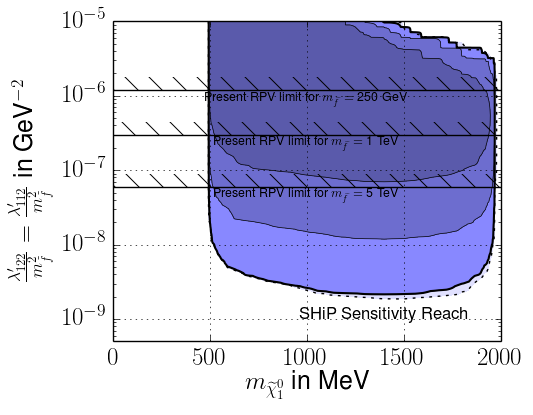} \qquad \qquad
\includegraphics[trim={0 0 0 0},clip,height=0.25\textheight]{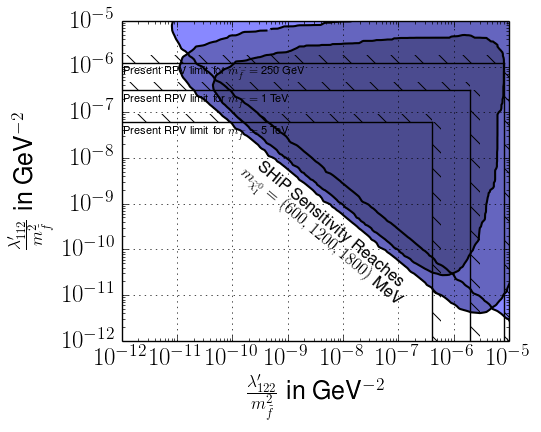} \\
\end{center}
\vspace{-1.25cm}
a) \hspace{0.45\textwidth} b) \hspace{0.4\textwidth} \hfill \\
\caption{Search sensitivity for Benchmark Scenario 2. The labelling is as in Fig.\,\ref{fig:121-112}, except for 
the couplings $\lam^\prime_{112},\,\lam^\prime_{122}$.}
\label{fig:112-122-Ds}
\end{figure*}

We see that the SHiP experiment can improve the current bounds on $\lam^\prime/m^2_{\tilde{f}}$ by one to three 
orders of magnitude depending on $m_{\tilde{f}}$. The kinematically accessible range for $m_{\tilde \chi^0_1}$ 
is dictated by the requirements that the neutralino must be lighter than the $D$-meson, yielding an upper bound, while 
at the same time being heavier than the $K$-meson, thus giving a lower bound. We find that the discovery region is 
mostly independent of $m_{\tilde \chi^0_1}$, \textit{i.e.} the lower solid curve edge of the discovery range is fairly flat,
as long as the mass lies within this kinematically allowed range between $500$ and $1900$\,MeV. 

The additional small lighter shaded region marked by the dashed line indicates the extended sensitivity if the SHiP detector 
could detect neutral kaons in the final state, as well. In this particular scenario, this barely affects the exclusion contours 
because the branching ratios to visible and invisible final states are roughly the same. However, as we see below, in 
other scenarios the difference can be more substantial.

If $\lam^\prime/m_{\tilde f}^2$ becomes larger than $10^{-5}\,\mathrm{GeV}^{-2}$, the neutralinos decay too fast,
in fact mostly before reaching the detector. We note that this parameter region is subject to large numerical 
uncertainties in our Monte Carlo simulation approach, and as such the exclusion lines show fluctuations with 
no underlying physical cause.
In Fig.~\ref{fig:121-112}b) we remove the restriction that the couplings $\lam^\prime_{121}$ and $\lambda^\prime_
{112}$ are equal. Instead, we present the SHiP sensitivity depending on the separate couplings for three representative 
values of $m_{\tilde \chi^0_1}$. We choose a light mass close to the lower kinematic threshold (600\,MeV, bright blue), 
a heavy one close to the higher kinematic threshold (1800\,MeV, dark blue) and one halfway between the other two 
(1200\,MeV, blue). In the same manner as before, existing limits on the two couplings are plotted for three representative 
values of $m_{\tilde{f}}$. Again we see that the contours are fairly insensitive to the neutralino mass. In all cases, SHiP 
probes a new region of parameter space even for $5$ TeV sfermion masses. The shape of the sensitivity regions 
are due to {\it both} couplings defining the neutralino lifetime, but only $\lam^\prime_{121}$ leads to the production 
and only $\lam^\prime_{112}$ to the \textit{observable} decay of the neutralinos. To avoid confusion, let us call the 
couplings $\lambda^\prime_P$ and $\lambda^\prime_D$ respectively in the following discussion. If $\lambda^\prime_P
/m_{\tilde{f}}^2$ becomes too small, too few neutralinos are produced in the first place. This leads to an overall minimum 
requirement on $\lambda^\prime_P/m_{\tilde{f}}^2$. 

For increasing $\lambda^\prime_P/m_{\tilde{f}}^2$, more neutralinos are produced and thus the allowed 
neutralino lifetime to observe 3 events at SHiP can be reached for increasingly larger ranges of $\lambda^
\prime_D/m_{\tilde{f}}^2$. As before, too small/large couplings lead to too many neutralinos decaying after/before 
the detector. Furthermore, smaller $\lambda^\prime_D/\lambda^\prime_P$ ratios lead to more invisibly decaying 
neutralinos. In Fig.~\ref{fig:121-112}b) this corresponds to the slanted edge running from the upper left-hand
corner to the lower right-hand corner.

Once $\lambda^\prime_P/m_{\tilde{f}}^2$ becomes too large, the lifetime induced by this operator is already too 
small and neutralinos decay mostly before reaching the detector, regardless of $\lambda^\prime_D/m_{\tilde{f}}^2$. 
This explains the sensitivity limitations on the right edge of Fig.~\ref{fig:121-112}b.

The analysis described here applies to the cases:
\begin{equation}
\mathrm{Production\!\!:}\, \lam^\prime_{i21},\quad \mathrm{Decay\!\!:} \,\lam^\prime_{j12},\quad i,j\not=3\,.
\end{equation}
Note that for $i=2$, neutralinos are produced via $D^\pm \rightarrow \widetilde{\chi}^0_1 + \mu^\pm$ and 
the extra muon would shift the upper kinematical limit of all regions in  Fig.~\ref{fig:121-112} by $m_\mu \approx 100\,$MeV 
to the left. Analogously, the case $j=2$ would move the lower kinematical limit of all but the shaded regions by the same 
amount to the right. Within this scenario, the cases $i=3$ and/or $j=3$ would not be observable as there would not 
be enough phase space to produce a $\tau$ lepton.

If SHiP is sensitive also to neutral final states, the above results also apply to the cases
\begin{equation}
\mathrm{Production\!\!:}\, \lam^\prime_{i21},\quad \mathrm{Decay\!\!:} \,\lam^\prime_{j21},\quad i,j\not=3\,.
\end{equation}
The sensitivity curve in this case would be very similar to the shaded region in Fig.~\ref{fig:121-112}a).

\subsection{Benchmark Scenario 2}

\begin{table}[thb]
\begin{tabular}{r||l}
\hline
$\lambda^\prime_P$ for production & $\lambda^\prime_{122}$ \\
$\lambda^\prime_D$ for decay & $\lambda^\prime_{112}$ \\
produced meson(s) & $D_s$ \\
visible final state(s) & $K^{\pm} e^\mp, K^{*\pm} e^\mp$ \\
invisible final state(s) via $\lambda^\prime_P$ & $(\eta, \eta^\prime, \phi) + (\nu, \bar{\nu})$ \\
invisible final state(s) via $\lambda^\prime_D$ & $(K^0_L,K^0_S, K^*) + (\nu, \bar{\nu})$ \\
\hline
\end{tabular}
\caption{Features of Benchmark Scenario 2.}
\label{tab:sec-2}
\end{table}

\begin{figure*}
\begin{center}
\includegraphics[trim={0 -0.5cm 0  0},clip,height=0.25\textheight]{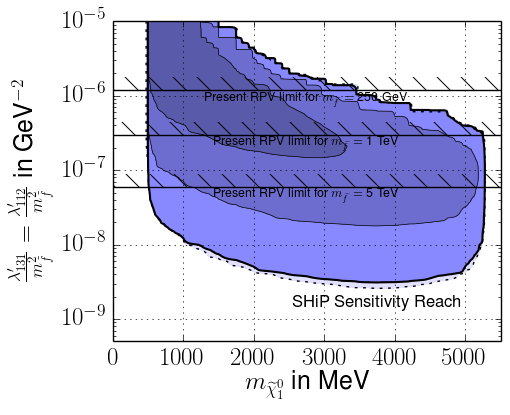} \qquad \qquad
\includegraphics[trim={0 0 0 0},clip,height=0.25\textheight]{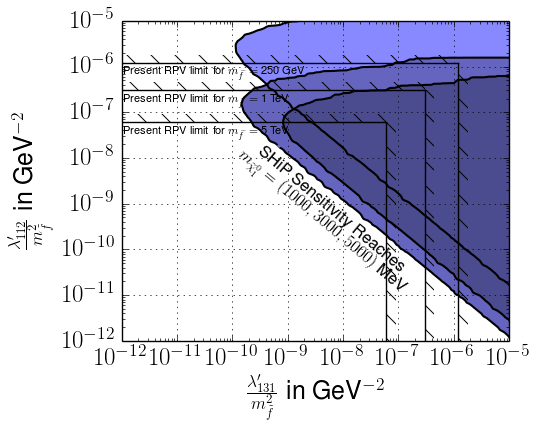} \\
\end{center}
\vspace{-1.25cm}
a) \hspace{0.45\textwidth} b) \hspace{0.4\textwidth} \hfill \\
\caption{Search sensitivity for Benchmark Scenario 3. The labelling is as in Fig.\,\ref{fig:121-112}, except for 
the couplings $\lam^\prime_{112},\,\lam^\prime_{131}$. Also in b) the neutralino masses are 1000\,MeV 
(light blue), 3000\,MeV (blue), and 5000\,MeV (dark blue). Note that the x-axis range for figure a) has been changed 
compared to previous cases, as the initial state $B$ meson allows for a larger kinematical reach.}
\label{fig:112-131-B}
\end{figure*}
This is similar to the previous benchmark, except the production of the neutralinos is via $D_s$
mesons. The observable charged final states are the same. There are however further invisible neutral final
states, which are kinematically accessible: $(\eta,\,\eta^\prime,\,\phi)+(\nu,\,\bar\nu)$. The details of this
benchmark scenario are summarized in Table\,\ref{tab:sec-2}. 

The results for the SHiP sensitivity are presented in Fig.\,\ref{fig:112-122-Ds}a). The reach is extended to higher 
neutralino masses, as $M_{D_s}>M_{D}$. The lower edge is still given by $m_{\tilde\chi^0_1}\approx M_K$. 
The sensitivity in $\lam^\prime/m_{\tilde f}^2$ is slightly weaker for two reasons. First, the production rate $D_s$ mesons is lower than $D^\pm$ mesons, see Table~\ref{tbl:prodnumerics}. Second, the neutralino branching ratio to an observable 
charged final state is smaller. Correspondingly the sensitivity is enhanced more in this scenario if neutral mesons 
are observable at SHiP, shown by the dashed line in the bottom of Fig.\,\ref{fig:112-122-Ds}b).

Analogously to Benchmark Scenario 1, the analysis described here applies to the cases
\begin{equation}
\mathrm{Production\!\!:}\; \lam^\prime_{i22},\quad \mathrm{Decay\!\!:} \;\lam^\prime_{j12},\quad i,j\not=3\,.
\end{equation}
If the neutral final-state mesons are visible, 
it also applies to the cases
\begin{align}
\mathrm{Production\!\!:}\; \lam^\prime_{i22},&\quad \mathrm{Decay\!\!:} \;\lam^\prime_{j21}, \lam^\prime_{j22},\quad i,j\not=3\,.
\end{align}

\subsection{Benchmark Scenario 3}

\begin{table}[H]
\begin{tabular}{r||l}
\hline
$\lambda^\prime_P$ for production & $\lambda^\prime_{131}$ \\
$\lambda^\prime_D$ for decay & $\lambda^\prime_{112}$ \\
produced meson(s) & $B^0, \bar{B}^0$ \\
visible final state(s) & $K^{\pm} e^\mp, K^{*\pm} e^\mp$ \\
invisible final state(s) via $\lambda^\prime_P$ & none \\
invisible final state(s) via $\lambda^\prime_D$ & $(K^0_L,K^0_S, K^*) + (\nu, \bar{\nu})$ \\
\hline
\end{tabular}
\caption{Features of Benchmark Scenario 3}
\label{tab:131-112-B}
\end{table}

In this scenario the neutralino production is via $B$ mesons. The decay of the neutralino via the coupling 
$\lam^\prime_{112}$ leads to charged $K$-mesons and electrons, which are readily visible. The coupling $\lam^\prime_
{112}$ also leads to neutral neutralino decays to $K^0$ mesons and neutrinos. There are no additional
kinematically accessible invisible neutralino decay modes through the coupling $\lam^\prime_{131}$. This 
information is summarized in Table~\ref{tab:131-112-B}.

The results of the simulation in this scenario are presented in Fig.\,\ref{fig:112-131-B}. The kinematically
accessible neutralino mass range is $M_{K^\pm} < m_{\tilde\chi^0_1}< M_{B^0}$. This is reflected in the 
shape of the sensitivity region in Fig.\,\ref{fig:112-131-B}a), which is cut off on the left at a neutralino mass 
of about 500\,MeV and on the right just under 5.3\,GeV. In the top-right corner, where $\lam^\prime/m^2_
{\tilde{f}}$ and $m_{\tilde\chi^0_1}$ are large, the neutralino lifetime becomes very short. The neutralinos 
then overwhelmingly decay before the detector. Since so few neutralinos reach the detector, we are here 
probing the extreme tail of the exponential decay distribution. Consequently, the top-right part of the
curve is jagged, due to lack of statistics in this regime. The lower curve slopes downward left to right, much 
more so than in Figs.\,\ref{fig:121-112}a) and \ref{fig:112-122-Ds}a). This effect is due to the presence of the 
final-state vector mesons ($K^*$), which are more important for the heavier neutralinos, accessible in 
$B$-meson decays. This is discussed in more detail in Section~\ref{subsec:vector}, below. The added sensitivity 
due to possible neutral final states is marginal, as the branching ratios are comparable.

As can be seen from the numbers in Table\,\ref{tbl:prodnumerics}, the $B$-meson production rate is roughly four
orders of magnitude smaller than the $D$-meson production rate. As the neutralino production is proportional to 
$(\lam^\prime/m^2_{\tilde{f}})^2$, the curves in Fig.\,\ref{fig:112-131-B}b) are shifted by almost two orders of 
magnitude to the right, compared to the corresponding results, \textit{e.g.} Fig.\,\ref{fig:121-112}b), of the previous 
benchmark scenarios. The new sensitivity reach of SHiP is thus smaller here. However, since $\lambda^\prime_
{131}$ does not induce any invisible decays of the neutralino, the sensitivity regions in Fig.\,\ref{fig:112-131-B}b) 
are not bounded on the right, in contrast to analogous regions of previous scenarios. Increasing $\lambda^\prime_
{131}/m_{\tilde{f}}^2$ then always leads to an increased number of expected neutralinos and hence always 
improves the sensitivity to $\lambda^\prime_{112}/m_{\tilde{f}}^2$.  We note that Fig.\,\ref{fig:112-131-B}b) has 
the same characteristic shape as Fig.\,6 in Ref.\,\cite{Dedes:2001zia}.

\begin{figure*}
\begin{center}
\includegraphics[trim={0 -0.5cm 0  0},clip,height=0.25\textheight]{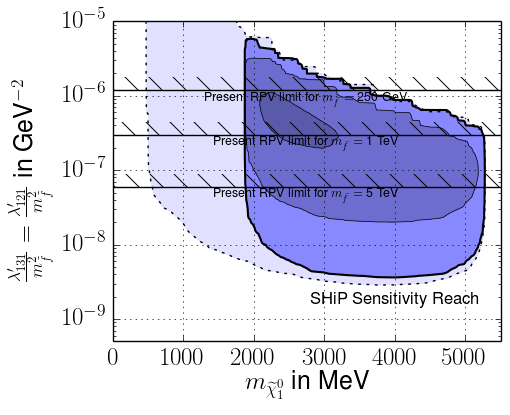} \qquad \qquad
\includegraphics[trim={0 0 0 0},clip,height=0.25\textheight]{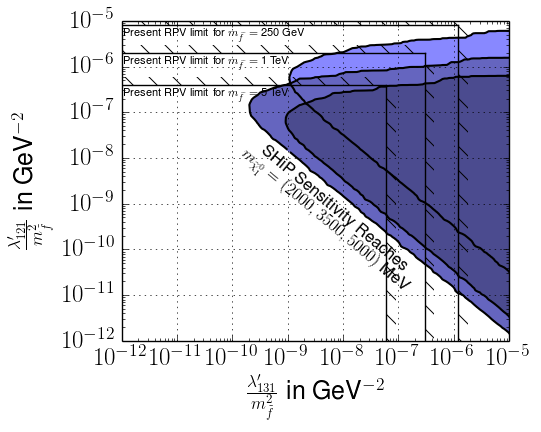} \\
\end{center}
\vspace{-1.25cm}
a) \hspace{0.45\textwidth} b) \hspace{0.4\textwidth} \hfill \\
\caption{Search sensitivity for Benchmark Scenario 4. The labelling is as in Fig.\,\ref{fig:121-112} except for 
the couplings $\lam^\prime_{131}$, $\lam^\prime_{121}$. Also in b) the neutralino masses are 2000\,MeV 
(light blue), 3500\,MeV (blue), and 5000\,MeV (dark blue).}
\label{fig:131-121-B}
\end{figure*}

The analysis described here applies to the cases
\begin{equation}
\mathrm{Production\!\!:}\; \lam^\prime_{i31},\,\lam^\prime_{i13},\quad \mathrm{Decay\!\!:} \;\lam^\prime_{j12},\quad i,j\not=3\,.
\end{equation}
Again, if neutral final state mesons can be observed it also applies to the cases
\begin{equation}
\mathrm{Production\!\!:}\; \lam^\prime_{i31},\,\lam^\prime_{i13},\quad \mathrm{Decay\!\!:} \;\lam^\prime_{j21},\quad i,j\not=3\,.
\end{equation}
Although as we discuss below in Section\,\ref{sub-sec:BS-4}, in this case there are additional charged decay 
modes via $D$-mesons.

\vfill
\subsection{Benchmark Scenario 4}
\label{sub-sec:BS-4}

\begin{table}[H]
\begin{tabular}{r||l}
\hline
$\lambda^\prime_P$ for production & $\lambda^\prime_{131}$ \\
$\lambda^\prime_D$ for decay & $\lambda^\prime_{121}$ \\
produced meson(s) & $B^0, \bar{B}^0$ \\
visible final state(s) & $D^{\pm} e^\mp, D^{*\pm} e^\mp$ \\
invisible final state(s) via $\lambda^\prime_P$ & none \\
invisible final state(s) via $\lambda^\prime_D$ & $(K^0_L,K^0_S, K^*) + (\nu, \bar{\nu})$ \\
\hline
\end{tabular}
\caption{Features of Benchmark Scenario 4}
\label{tab:131-121-B}
\end{table}

In this scenario the neutralinos are also produced via $B$-mesons and the coupling $\lam^\prime_{131}$.
However, now the decay is into $D$ mesons via the coupling $\lam^\prime_{121}$. There are kinematically
accessible invisible decays to neutral $K$ mesons and neutrinos via $\lam^\prime_{121}$ This is summarized in 
Table\,\ref{tab:131-121-B}.

The results of the simulation for this scenario are displayed in Fig.\,\ref{fig:131-121-B}. In the left panel, we show the 
search sensitivity as a function of the neutralino mass and of a common coupling $\lam^\prime_{113}/m_{\tilde f}^2=
\lam^\prime_{112}/m_{\tilde f}^2\equiv\lam^\prime/m_{\tilde f}^2$. The mass sensitivity range is again mainly fixed 
kinematically: $M_D<m_{\tilde\chi^0_1}<M_B$, which is narrower than in Fig.\,\ref{fig:112-131-B} due to the larger 
$D$-meson mass. 

When allowing for neutral final states the sensitivity is dramatically increased to lower neutralino masses, 
corresponding to the kinematic range $M_K<m_{\tilde\chi^0_1}<M_B$. This is shown in  Fig.\,\ref{fig:131-121-B}a) 
by the very light blue region bounded by the dashed line.

In Fig.\,\ref{fig:131-121-B}b) the sensitivity range is similar to Fig.\,\ref{fig:112-131-B}b), as it is dominated by 
$B$-meson production. The differences in the curves are mainly due to the different neutralino masses that are 
considered: 2000\,MeV (light blue), 3500\,MeV (blue), and 5000\,MeV (dark blue).

\begin{figure*}
\begin{center}
\includegraphics[trim={0 -0.5cm 0  0},clip,height=0.25\textheight]{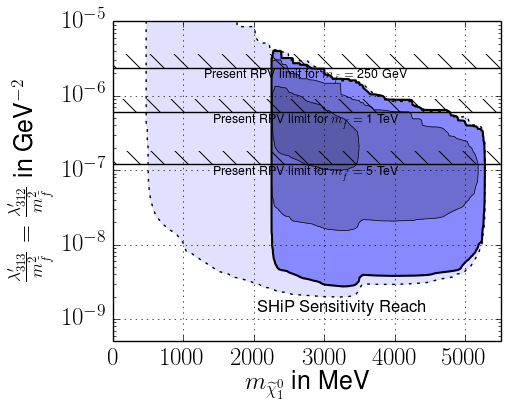} \qquad \qquad
\includegraphics[trim={0 0 0 0},clip,height=0.25\textheight]{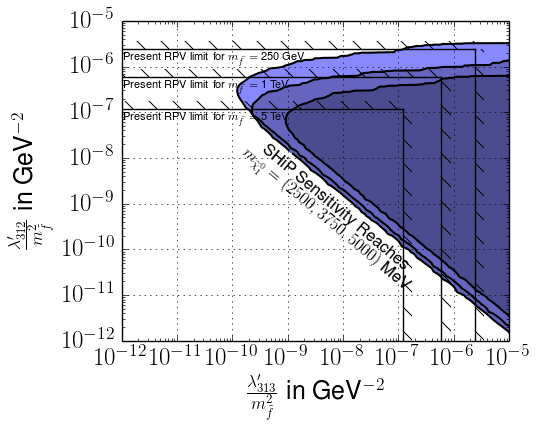} \\
\end{center}
\vspace{-1.25cm}
a) \hspace{0.45\textwidth} b) \hspace{0.4\textwidth} \hfill \\
\caption{Search sensitivity for Benchmark Scenario 5. The labelling is as in Fig.\,\ref{fig:121-112} except for 
the couplings $\lam^\prime_{313}$, $\lam^\prime_{312}$. Also in b) the neutralino masses are 2750\,MeV 
(light blue), 3750\,MeV (blue), and 5000\,MeV (dark blue).}
\label{fig:313-312-B}
\end{figure*}

The analysis described here also applies to the cases
\begin{equation}
\mathrm{Production\!\!:}\; \lam^\prime_{i31},\,\lam^\prime_{i13},\quad \mathrm{Decay\!\!:} \;\lam^\prime_{j21}\quad i,j\not=3\,.
\end{equation}

\subsection{Benchmark Scenario 5}

\begin{table}[H]
\begin{tabular}{r||l}
\hline
$\lambda^\prime_P$ for production & $\lambda^\prime_{313}$ \\
$\lambda^\prime_D$ for decay & $\lambda^\prime_{312}$ \\
produced meson(s) & $B^0, \bar{B}^0, B^\pm (+\,\tau^\mp)$  \\
visible final state(s) & $K^{\pm} \tau^\mp, K^{*\pm} \tau^\mp$ \\
invisible final state(s) via $\lambda^\prime_P$ & none \\[1mm]
invisible final state(s) via $\lambda^\prime_D$ & $(K^0_L,K^0_S, K^*) + (\nu, \bar{\nu})$ \\
\hline
\end{tabular}
\caption{Features of Benchmark Scenario 5. At the end of the third row
we emphasize that the charged $B$-meson decay to the neutralino is accompanied by a tau.}
\label{tab:313-312-B}
\end{table}

Here the production goes via $B$-mesons and the coupling $\lam^\prime_{313}$. We thus consider the third 
generation lepton index. The features of this benchmark scenario are summarized in Table\,\ref{fig:313-312-B}. At 
the end of the third row, the $(+\,\tau^\pm)$ indicates that the charged $B$-meson decay to a neutralino is 
accompanied by a tau,
\begin{equation}
B^\pm\to \tilde\chi^0_1 + \tau^\pm\,.
\end{equation}
Therefore the charged $B^\pm$ meson can only contribute for the restricted neutralino mass range $m_{\tilde\chi
^0_1}< M_B-m_\tau$. The corresponding {\it neutral} $B$-meson decay has neutrinos in the final state and is thus 
allowed for the larger range $m_{\tilde\chi^0_1}< M_B$. In Fig.\,\ref{fig:313-312-B}a) this leads to a kink in the 
solid curves surrounding the sensitivity regions at $m_{\tilde\chi^0_1}= M_B-m_\tau$. The region corresponding to 
$\geq10^6$ events is also cut off to higher neutralino masses compared to Fig.\,\ref{fig:131-121-B}a). 

The neutralino decay proceeds via the coupling $\lam^\prime_{312}$. 
\begin{equation}
\tilde\chi^0_1 \to K^\pm+ \tau^\mp\,.
\end{equation}
Thus the visible final states involve charged $K$-mesons {\it and} tau leptons, which might be difficult to 
detect, especially in the hadronic decay mode. This also requires $m_{\tilde\chi^0_1}>M_K+m_\tau\simeq
2300\,$MeV. This is the cutoff on the left of the blue regions in Fig.\,\ref{fig:313-312-B}a).

There are possible additional neutral final states  involving neutral $K$-mesons and neutrinos and {\it no} 
tau lepton. When these are included the sensitivity reach is dramatically extended to lower neutralino masses 
as can be seen in Fig.\,\ref{fig:313-312-B}a), just as in Fig.\,\ref{fig:131-121-B}a).

Fig.\,\ref{fig:313-312-B}b) shows the sensitivity region as a function of the two now 
independent couplings $\lam^\prime_{313}/m_{\tilde f}^2$ and $\lam^\prime_{312}/m_{\tilde f}^2$ for the three neutralino
masses 2750\,MeV (light blue), 3750\,MeV (blue), and 5000\,MeV (dark blue). These are slightly modified 
compared to Scenario 4, because of the tau mass, leading to slightly different curves.

The analysis described here applies to the cases
\begin{equation}
\mathrm{Production\!\!:}\; \lam^\prime_{i13},\,\lam^\prime_{i31},\quad \mathrm{Decay\!\!:} \;\lam^\prime_{j12}\,
\end{equation}
with either $i$ or $j$ or both equal to $3$. In case of $\lambda^\prime_{i31}$ only neutral $B$-mesons can 
decay into a neutralino and corresponding neutrino such that the sensitivity curves are independent of $i$. 
The curves look similar to those in Fig.~\ref{fig:313-312-B}, but the kink around $M_B-m_\tau\simeq$ 
3500\,MeV does not appear. The case $j = 1$ leads to similar limits as the dashed region drawn in 
Fig.\,\ref{fig:313-312-B}b), whereas similarly to previous scenarios it is shifted by $m_\mu$ to the right for $j = 2$.

\subsection{Related Scenarios}

As we have seen the sensitivity curves are largely shaped by the kinematics. There are thus some related
scenarios which we briefly describe here. Benchmark scenarios 3, 4 and 5 are easily extended  to the cases 
\begin{equation}
\mathrm{Production\!\!:}\; \lam^\prime_{i32},\,\lam^\prime_{i23},
\end{equation}
by considering the production via $B_s$ mesons. As can be seen from  Table\,\ref{tbl:prodnumerics}, $B_s$ 
production is suppressed compared to $B$-meson production by roughly a factor $n_{B^0}^{b\bar b}/n_{B^0_s}
^{b\bar b}\approx0.18$. The sensitivity to the coupling $\lambda^\prime/m^2_{\tilde{f}}$ is reduced by
roughly a factor $\sqrt[4]{0.18}\approx0.65$.

Similarly the decay in Scenarios 4 and 5 are easily extended to the cases $\lam^\prime_D=\lam^\prime_{122}$, 
where $D_s$ mesons appear in the final state. This gives rise to very similar sensitivity curves, the only difference 
being a slightly larger lower-mass reach for the neutralino. 

\subsection{Relevance of Vector Mesons}
\label{subsec:vector}

\begin{figure}
\begin{center}
\includegraphics[trim={0 0 0 0},clip,height=0.25\textheight]{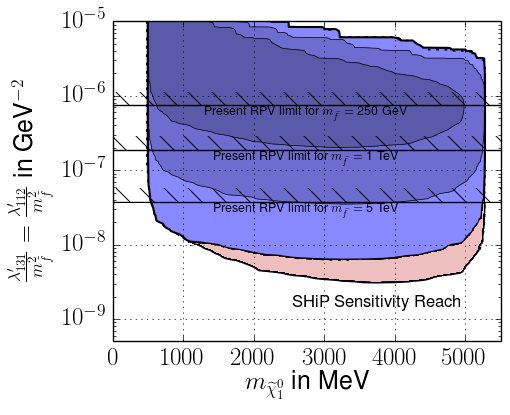} \\
\end{center}
\caption{Sensitivity curve for enchmark Scenario 3 if decays into vector mesons are ignored. The pink region shows the sensitivity curve including vector mesons which corresponds to the ``SHiP sensitivity reach'' of Fig.~\ref{fig:112-131-B}a)}.
\label{vectormesons}
\end{figure}
As discussed at the end of Sec.~\ref{sec:rpvinteractions:formalism}, the inclusion of vector mesons in the final state leads to a reduced sensitivity to potential fine-tuning of the SUSY parameters. 
As discussed at the end of Sec.~\ref{sec:rpvinteractions:formalism}, the inclusion of vector mesons in the final state lead to complementary dependence on SUSY paramaters. In addition, the final 
state vector mesons lead to an interesting kinematical enhancement of the neutralino decay width. This 
enhancement can be understood from the decay width formul{\ae} in Eqs.~\eqref{eq:width3} and \eqref{eq:width4}. 
When $m_{\tilde\chi^0_1}\gg m_M$, where $m_M$ is the mass of the final state meson, the decay to scalar mesons 
is proportional to $(f^{S}_M)^2 m^2_{\tilde\chi^0_1}\simeq f_M^2 m_M^2 m^2_{\tilde\chi^0_1} $, whereas the decay 
to vector mesons is proportional to $(f^{T}_M)^2 m^4_{\tilde\chi^0_1} \simeq f_M^2 m^4_{\tilde\chi^0_1}$. The decay 
into vector mesons is thus enhanced by roughly a factor $m^2_{\tilde\chi^0_1}/m_M^2$. 

This is mainly relevant for cases such as our Benchmark Scenario 3, where neutralino production occurs via $B$-meson 
decays. Then the neutralino can be significantly heavier than the final-state meson, here the kaon. To illustrate the 
enhancement, in Fig.~\ref{vectormesons}, we repeat the analysis for Benchmark Scenario 3, but exclude 
final state vector mesons. The sensitivity contour of the same analysis including vector mesons is shown by the pink shaded area. This region is identical to the sensitivity area shown in Fig.~\ref{fig:131-121-B}a). We show it again here to make the difference between the two cases easier to see. The inclusion of vector mesons is barely visible for neutralino masses smaller than $1$ GeV, 
but the enhancement is clearly visible in the 2-5 GeV mass range.

\section{LHC Estimate}
\label{sec:lhc-estimate}
\begin{figure}
\vspace{-0.5cm}
\def\svgwidth{250pt} 
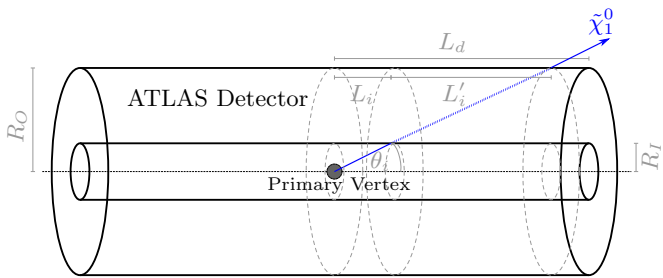
\vspace{-0.5cm}
\caption{Schematic overview of the ATLAS detector geometry and definition of distances and angles used in text.}
\label{fig:geometrylhc}
\end{figure}
As we have seen, the sensitivity at SHiP in the production coupling results from an interplay between the length scales of the target-
detector-distance $L_{t \rightarrow d}$, the length of the detector $L_d$, the meson production rate $N_M$, the boost $\gamma_i$ 
of the neutralinos and their azimuthal angle $\theta_i$. The LHC operates at much higher energies in the center-of-mass frame and 
the detectors are built right at the collision points. Thus all the above parameters change and we would expect a different sensitivity 
when comparing to SHiP\footnote{We thank Jesse Thaler for drawing our attention to this point. See also \cite{Ilten:2015hya} on
a related discussion on dark photons.}\!\!.  To estimate the net result of these effects, we thus here briefly discuss the sensitivity for 
our scenarios at the LHC.

To allow for an easy comparison, we consider two example cases which correspond to our earlier discussed Benchmark Scenarios 
1 and 5. These involve the observable decay chains
\begin{align}
\hspace{-0.2cm} D^\pm \to \tilde\chi^0_1 e^\pm, &\quad \tilde\chi^0_1\to e^\pm K^\mp\, \quad \text{via $\lambda^\prime_{121},
\lambda^\prime_{112}$}, \\
\hspace{-0.2cm}B^\pm/B^0 \to \tilde\chi^0_1 \tau^\pm/\nu, &\quad \tilde\chi^0_1\to \tau^\pm K^\mp\, \quad \text{via $\lambda^\prime_{313},
\lambda^\prime_{312}$}.
\end{align}

To compare like with like, we estimate the neutralino event rates analogously to Sec.~\ref{sec:simulation}: we simulate these 
scenarios using \verb@Pythia 8.175@ \cite{Sjostrand:2006za, Sjostrand:2007gs}, and find a production cross section for 
$c\bar c$ at 14 TeV of $6 \cdot10^{12}\,$fb and $\sigma_{b\bar{b}}/\sigma_{c\bar{c}} = 8.6 \times 10^{-3}$. We consider an integrated luminosity of 250\,fb$^{-1}$, which roughly corresponds 
to the expected value for a high-energy LHC running for 5 years. We determine the other parameters of interest as in 
Sec.~\ref{sec:simulation} and list them in Table~\ref{tbl:prodnumericslhc}.

As an example we consider the ATLAS detector setup as sketched in Fig.~\ref{fig:geometrylhc}. Here we assume the detectable 
region to approximately range from $R_I=0.0505\,$m, the beginning of the inner detector, to $R_O=11$~m, the end of the muon 
chambers. The detector has cylindrical shape with a total length of $ 2 L_D =  43$~m. The probability for the neutralino to decay within this range is then, similarly to Eq.~(\ref{eq:observechi}), 
\begin{align}
P[(\tilde\chi^0_{1})_i \text{ in d.r.}] &= e^{-L_i/\lambda^z_i} \cdot \Big(1 - e^{-L_i^\prime/\lambda^z_i}\Big), \label{eq:observechilhc} \\
L_i &\equiv \text{min}(L_d, |R_I/\tan \theta_i|) \\
L_i^\prime &\equiv \text{min}(L_d, |R_O/\tan \theta_i|) - L_i
\end{align}
with angles and distances defined in Fig.~\ref{fig:geometrylhc} and $\lambda^z_i$ as defined in Eq.~(\ref{eq:probdefinitions1}). In similar 
manner as for SHiP, we use \texttt{Pythia} to simulate $20,\!000$ events, force all mesons of the right type to decay into neutralinos and 
average the results for $P[(\tilde\chi^0_{1})_i \text{ in d.r.}]$ over all these Monte-Carlo neutralinos to find the overall probability that an 
LHC-produced neutralino decays inside the detectable region of the ATLAS detector. The number of observable neutralino decays is then 
determined by considering the RPV branching ratio of the initially produced mesons and the branching ratio of neutralinos into charged 
final states, according to Eqs.~(\ref{eq:chiprod}), (\ref{eq:chidec}).

\begin{table}
\begin{tabular}{lcl}
\toprule
$N_{c\bar{c}}$ & \quad & $1.5 \times 10^{15}$ \\[0.6mm]
$\sigma_{b\bar{b}} / \sigma_{c\bar{c}}$ & & $8.6 \times 10^{-3}$ \\ [0.9mm]
\hline
$ n^{c\bar c }_{D^\pm} $ & & 0.59 \\ [0.8mm]
\hline
$ n^{b\bar b }_{B^\pm}$ & & 0.87 \\ [0.8mm]
$ n^{b\bar b }_{B^0}$ & & 0.87 \\[0.8mm]
\botrule
\end{tabular}
\caption{Numerical values used to estimate the number $N_M$ in Eq.~(\ref{eq:prodmes}) for the LHC with $\sqrt{s} = 14~$TeV and an integrated luminosity of 250 fb$^{-1}$. All numbers are evaluated by simulating 1M events of each 
\texttt{HardQCD} type in \texttt{Pythia}. }
\label{tbl:prodnumericslhc}
\end{table}

\begin{figure*}
\begin{center}
\includegraphics[trim={0 -0.5cm 0  0},clip,height=0.25\textheight]{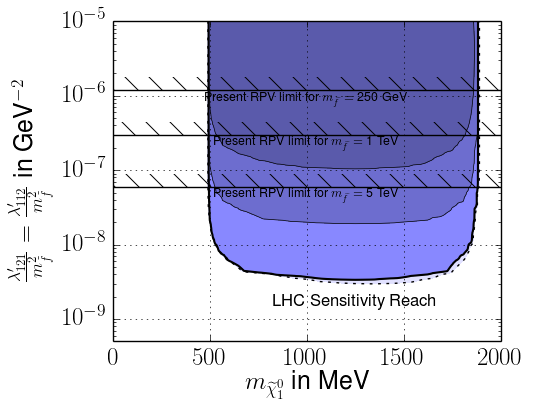} \qquad \qquad
\includegraphics[trim={0 -0.5cm 0  0},clip,height=0.25\textheight]{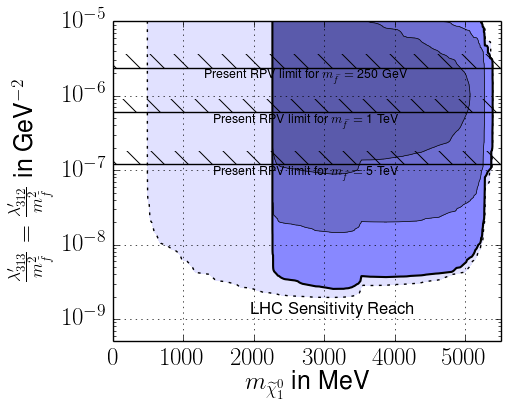}
\end{center}
\vspace{-1.25cm}
a) \hspace{0.45\textwidth} b) \hspace{0.4\textwidth} \hfill \\
\caption{Expected LHC search sensitivity for a) Benchmark Scenario 1 and b) Benchmark Scenario 5. The labelling is as in 
Figs.\,\ref{fig:121-112},\ref{fig:313-312-B}.}
\label{fig:131-121-Blhc}
\end{figure*}

The expected sensitivity regions are shown in Fig.~\ref{fig:131-121-Blhc}.  For easy comparison we show the same information 
as in their corresponding SHiPs analogues Figs.~\ref{fig:131-121-B}a) and \ref{fig:313-312-B}a) and focus on the 3 event 
threshold which would be required for a significant observation. We restrict the discussion to the $m_{\tilde{\chi}_1^0}$--$\lambda
^\prime/m^2_{\tilde{f}}$--plane as the limits in the $\lambda^\prime_{\text{prod.}}/m^2_{\tilde{f}}$--$\lambda^\prime_{\text{dec.}}/
m^2_{\tilde{f}}$--plane can be easily deduced.

For both scenarios, the structure of the plots do not largely differ between ATLAS and SHiP. The testable kinematic regions 
are obviously identical for the same scenario and it is only the required value for $\lambda^\prime/m^2_{\tilde{f}}$ to observe 
enough neutralino decays which changes. At SHiP we found that the expected sensitivity quickly drops if the neutralinos decay 
too promptly, that is before they reach the decay chamber at roughly 70~m behind the target. This resulted in an upper limit on 
the couplings SHiP would be sensitive to. However, as the detectable region at ATLAS already starts at $\mathcal{O}(\text{cm})$ 
distances from the primary vertex, this upper limit is pushed to higher values. 
\begin{figure*}
\includegraphics[width=0.8\columnwidth]{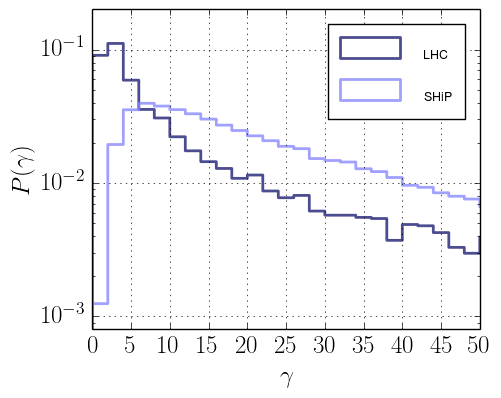} \qquad
\includegraphics[width=0.8\columnwidth]{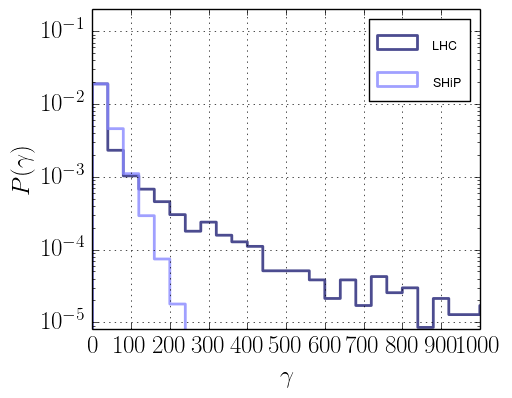}
\caption{Boost distribution for neutralinos produced at a $\sqrt{s}=14$~TeV LHC and at SHiP, determined with $\texttt{Pythia}$. We show the same results for two different choices of axis scaling.  The fluctuations for $\gamma \gtrsim 300$ are caused by limited Monte Carlo statistics. }
\label{fig:boost}
\end{figure*}

Comparing the results for Scenario 1, Figs.~\ref{fig:131-121-B}a) and \ref{fig:131-121-Blhc}a), we find that LHC has a comparable 
but still by a factor 2 weaker expected sensitivity on $\lambda^\prime/m^2_{\tilde{f}}$ than the expected value from SHiP. From 
comparing the respective values for $N_{c\bar{c}}$ in Tables \ref{tbl:prodnumericslhc} and \ref{tbl:prodnumericslhc}, one expects 
SHiP to produce almost 100 times more neutralinos than the LHC in a comparable time frame. This is partially compensated by the 
effect that neutralinos which are produced at large angles $\theta$ can be observed at the almost spherical ATLAS detector but 
miss the decay chamber at SHiP. 

Furthermore, the boost distribution of the two experiments largely differ. The fixed target setup of SHiP causes most produced mesons to have a large boost which is inherited by the daughter neutralinos they decay into. Contrarily, the center-of-mass collision for the LHC will lead to most mesons to be produced at rest. We show the boost distribution of the neutralinos we get with \texttt{Pythia} in Fig.~\ref{fig:boost}. For SHiP, the distribution shows an  expectation value of  
$\langle \gamma \rangle \approx 30$ and a maximum probability for $\gamma_{\text{max}} \approx 7.5$. This leads to an increased 
lifetime in the lab frame which reduces the detection probability if $c \tau_\chi$ is larger than the size of the detector, see \textit{e.g.} 
Eq.~(\ref{eq:observechi}). 

The large center-of-mass energy of the LHC leads to an even larger {\it average} boost, $\langle \gamma \rangle \approx 55$, 
which however has a larger spread, resulting in many neutralinos with boost of $\mathcal{O}(1)$ and a few with boost 
$\mathcal{O}(1000)$. As shown in Fig.~\ref{fig:boost}, the peak of the boost distribution for the LHC is located at $\gamma = 2.5$ and the resulting large fraction of unboosted neutralinos improve the overall probability to observe a decay within 
ATLAS.  

Combining the above effects, we find that $\langle P[\tilde\chi^0_{1} \text{ in d.r.}]\rangle$ at ATLAS is greater
by a factor 4 than at SHiP. Taking into account the much larger meson production yield of SHiP, it is expected to observe 
approximately 25 times more events than ATLAS, leading to an improved sensitivity on the coupling of $\sqrt[4]{25} \approx 2.2$. 

For Scenario 5, we compare Figs.~\ref{fig:313-312-B}a) and \ref{fig:131-121-Blhc}b) and interestingly find very comparable 
expected sensitivities. All the effects discussed for the previous scenario equally apply here and lead to approximately similar 
results. Therefore one would still expect SHiP to observe roughly 25 times more decays. However, this scenario requires 
$B$-mesons to be initially produced. The larger center-of-mass energy at LHC leads to an increased relative production yield 
$\sigma_{b\bar{b}} / \sigma_{c\bar{c}}$ of approximately 40 (see Tables \ref{tbl:prodnumerics}, \ref{tbl:prodnumericslhc}). This 
results in a roughly \unit[60]{\%} larger overall expected event rate at the LHC, which however is a negligible 
improvement when translated into a limit on $\lambda^\prime/m^2_{\tilde{f}}$.

It is clear that the results we show can just serve as a very approximate comparison. As we do not know the efficiency with 
which SHiP would be able to detect a neutralino decay and distinguish it from Standard Model, we did not take it into account 
for our LHC discussion either. However, it can be expected that the final state efficiencies for the two experiments  differ 
significantly, most likely with a significant penalty on the ATLAS side. SHiP will be specifically designed to observe rare decays 
of new, long-lived particles. It can therefore be expected that the neutralino decays will have a large 
probability to actually be measured by the detector. The ATLAS detector, however, is not designed for this purpose. The 
combined efficiencies to trigger on the event, to reconstruct the final state particles, to identify the significantly displaced vertex 
and to distinguish it from Standard Model mesons decays will most likely lead to a significant reduction of the final event yield, potentially by orders of magnitude. 
Still, we find it an interesting observation that when just considering the 
geometry of the setup, the meson production yield and the expected kinematics of the neutralinos, SHiP and ATLAS seem to have 
comparable sensitivity to the discussed decay scenario. Thus we conclude that the final state reconstruction efficiency will play 
a crucial role in determining the importance of the SHiP experiment with regards to the search for light neutralinos.

\section{Summary}
\label{sec:conclusion}

In this work we have studied the sensitivity of the proposed SHiP experiment to the R-parity violating production and 
decay of neutralinos whose masses lie in the range of $0.5-5$ GeV. As discussed in Sect.~\ref{sec:light-neutralino}, a 
neutralino in this mass range is only allowed if R-parity is violated. We have focused on the semi-leptonic R-parity 
violating operators $\lambda^\prime LQ\bar D$, but our work is easily extended to the purely leptonic case $\lambda 
LL\bar E$. The basic idea pursued in this work is that a small fraction of the $D$ and $B$ mesons produced in the 
SHiP experiment, can potentially decay into a neutralino plus lepton. Because the R-parity violating couplings are 
expected to be small, the neutralino can have a sufficiently long lifetime to travel a distance of 63.8\,m to the ShiP 
detector where it can subsequently decay into a, presumably detectable, meson-lepton pair.

For neutralinos in the mass-range of $0.5-5$ GeV, the SHiP experiment is sensitive to several combinations of R-parity 
violating couplings. In general, the number of neutralino decays in the SHiP detector is proportional to 
$(\lambda^\prime
_{iab} \lambda^\prime_{jcd})^2/m^8_{\tilde f}$, where $i,j$ denote the lepton generation indices and $a,b,c,d$ the quark 
generation indices. We have classified benchmark scenarios for different combinations of the generation indices $i,j,a,b,c,d$. 
Although many different combinations of couplings exist, we have argued that most of them can be captured in this 
relatively small set of benchmark scenarios.
We highlight here a number of conclusions and caveats of our findings. 
\begin{itemize}
\item We find no feasible scenario where SHiP is sensitive to only a single $\lambda^\prime_{iab}$ coupling. The main 
obstacle for such a scenario is that the final state decay products will consist of a neutrino and a neutral meson from 
which it is difficult to reconstruct the neutralino decay.  An example of such a scenario would be a nonzero $\lambda^
\prime_{i21}$ coupling, which would lead to the production and decay channels $D^\pm \rightarrow \tilde \chi^0_1 + 
l_i^\pm$ and $\tilde \chi^0_1\rightarrow K_{S,L}+\nu$. In order to get an observable final state we thus always require two 
distinct nonzero $\lambda^\prime$ couplings. We have found, however, that including the invisible final states is 
mandatory as they influence the neutralino lifetime. 

\item That being said, we have shown that the SHiP experiment has the potential to significantly improve the constraints 
on various combinations of R-parity violating couplings. This is clearly illustrated in the figures in Sect.~\ref{sec:results}. 
For instance, constraints on $\lambda^\prime_{112}/m^2_{\tilde f}$ can be 
strengthened by one to three orders of magnitude (see Fig.~\ref{fig:121-112}), depending on the sfermion mass. Similar 
improvements are found for third generation couplings such as $\lambda^\prime_{131} \lambda^\prime_{121}/m^4_
{\tilde f}$. We have presented sensitivity curves for each of the benchmark scenarios: $[\lam'_{121},\,\lam'_{112}],
\;[\lam'_{122},\,\lam'_{112}],\;[\lam'_{131},\,\lam'_{112}],\;[\lam'_{131},\,\lam'_{121}],$ $[\lam'_{313},\,\lam'_{312}]$. These 
curves can be used to estimate the sensitivity of the SHiP experiment to various combinations of R-parity violating 
interactions, as outlined in the text.

\item We have focused on neutralino production via the decay of $B$ and $D$ mesons. Very light neutralinos could be 
produced in kaon decays and be detected by their subsequent decay into pionic final states. We have not included this 
in this work as the production of light mesons is not well simulated in the forward direction with \texttt{Pythia}. We aim to 
study this in future work as it would extend the sensitivity to the neutralino mass range $0.1-0.5$ GeV. Similarly, we plan 
to include neutralino production in the decays of $\bar b b$ mesons, which would give a sensitivity to $\lambda^\prime_
{i33}$ and to higher-mass neutralinos.

\item We have found that including vector mesons in the final state leads to an enhanced sensitivity to neutralinos at the 
higher end of the allowed mass range. This enhancement can be understood from the decay-width formula presented in 
Sect.~\ref{sec:decay} as discussed in Sect.~\ref{subsec:vector}. In addition, the neutralino decay into vector mesons is 
proportional to a different combination of SUSY parameters than the corresponding decay into scalar mesons. The 
processes are therefore complementary.

\item Our analysis did not include possible uncertainties arising from hadronic matrix elements. The decay constants used 
are not in all cases known to high precision. Nevertheless, the SHiP sensitivity curves range over many orders of magnitude 
and we do not expect that changes in the decay constants drastically change our conclusions. 

\item We determined the expected sensitivity of the ATLAS detector at a \unit[14]{TeV} LHC with an integrated luminosity 
of \unit[250]{fb$^{-1}$}. To do a fair comparison, we did not take into account the final state reconstruction efficiency and 
only determined the expected number of neutralino decays inside the detector region of ATLAS. We found that in scenarios with initially produced $D$ mesons, ATLAS expects roughly \unit[4]{\%} of the number of events expected for SHiP, leading to an expected 
limit on $\lambda^\prime/m^2_{\tilde{f}}$ which is weaker by roughly a factor of 2. This is caused by a combination of a larger 
meson flux expected for SHiP and a higher detection probability for long-lived neutralinos at ATLAS. For initially produced $B$ 
mesons, the expected sensitivities are very similar, as the large LHC energies will produce relatively more $b$-quarks. It is 
therefore the final state reconstruction efficiency which will be the decisive factor.
\end{itemize}

\subsection*{Note added}
While completing this work, a related study appeared \cite{Gorbunov:2015mba}. They consider the production via 
$B$-mesons as in \cite{Dedes:2001zia} and the decay to kaons or purely leptonically. They make  extensive use 
of the formul{\ae} in the original study as presented by two of us (HKD and DS) in \cite{Alekhin:2015byh}, and thus
employs the error we made there, as discussed here in footnote \ref{footnote1}.

\acknowledgments
This work (JdV) is supported in part by the DFG and the NSFC through funds provided to the Sino-German CRC 110 
``Symmetries and the Emergence of Structure in QCD'' (Grant No. 11261130311). We thank Christoph Hanhart for 
several useful discussions. We thank Jesse Thaler for drawing our attention to the possibility of searching for these 
scenarios at the LHC. One of us (HD) thanks the Galileo Galilei Institut in Florence for hospitality, where a significant 
part of this work was completed.

\bibliography{RPV-Light-Neutralino}

\end{document}